\documentclass[twocolumn]{aastex61}
\usepackage{amssymb}
\usepackage{multirow}
\usepackage{morefloats}
\usepackage{amsmath}
\usepackage{bigstrut}
\usepackage{booktabs}
\usepackage{natbib}
\usepackage{float}
\usepackage{graphicx,epsfig,fancyhdr,epsf,txfonts,epstopdf}
\usepackage{latexsym,bbm}
\usepackage{lineno}
\usepackage{color}
\usepackage{ulem}
\usepackage{stfloats}

\received{XXX}
\revised{YYY}
\accepted{ZZZ}

\shortauthors{Gupta et al.}

\begin{document}

\title{Characterizing optical variability of OJ 287 in 2016--2017}

\author{Alok C.\ Gupta}\thanks{Email: acgupta30@gmail.com}
\affil{Aryabhatta Research Institute of Observational Sciences (ARIES), Manora Peak, Nainital -- 263002, India} 
\affiliation{Key Laboratory for Research in Galaxies and Cosmology, Shanghai Astronomical Observatory, Chinese Academy
of Sciences, Shanghai 200030, China}

\author{Haritma Gaur}\thanks{Email: harry.gaur31@gmail.com}
\affiliation{Aryabhatta Research Institute of Observational Sciences (ARIES), Manora Peak, Nainital -- 263002, India} 

\author{Paul J. Wiita}\thanks{E-mail: wiitap@tcnj.edu}
\affiliation{Department of Physics, The College of New Jersey, 2000 Pennington Rd., Ewing, NJ 08628-0718, USA}

\author{A. Pandey}
\affiliation{Aryabhatta Research Institute of Observational Sciences (ARIES), Manora Peak, Nainital -- 263002, India}

\author{P. Kushwaha}
\affiliation{Department of Astronomy (IAG-USP), University of Sao Paulo, Sao Paulo 05508-090, Brazil}

\author{S. M. Hu}
\affiliation{Shandong Provincial Key Laboratory of Optical Astronomy and Solar-Terrestrial Environment, Institute of Space 
Sciences, Shandong University, Weihai 264209, China}

\author{O. M. Kurtanidze}
\affiliation{Abastumani Observatory, Mt. Kanobili, 0301 Abastumani, Georgia} 
\affiliation{Engelhardt Astronomical Observatory, Kazan Federal University, Tatarstan, Russia} 
\affiliation{Center for Astrophysics, Guangzhou University, Guangzhou 510006, China} 
\affiliation{Key Laboratory of Optical Astronomy, National Astronomical Observatories, Chinese Academy of Sciences, 
Beijing 100012, China}

\author{E. Semkov}
\affiliation{Institute of Astronomy and National Astronomical Observatory, Bulgarian Academy of Sciences, 
72 Tsarigradsko Shosse Blvd., 1784 Sofia, Bulgaria}

\author{G. Damljanovic}
\affiliation{Astronomical Observatory, Volgina 7, 11060 Belgrade, Serbia}

\author{A. Goyal}
\affiliation{Astronomical Observatory of Jagiellonian University, ul. Orla 171, 30-244 Kraków, Poland}

\author{M. Uemura}
\affiliation{Hiroshima Astrophysical Science Center, Hiroshima University, Kagamiyama 1-3-1, Higashi-Hiroshima
739-8526, Japan}

\author{A. Darriba}
\affiliation{American Association of Variable Star Observers (AAVSO), 49 Bay State Rd., Cambridge, MA 02138, USA}
\affiliation{Group M1, Centro Astron$\acute{o}$mico de Avila, Madrid, Spain}

\author{Xu Chen}
\affiliation{Shandong Provincial Key Laboratory of Optical Astronomy and Solar-Terrestrial Environment, Institute of Space
Sciences, Shandong University, Weihai 264209, China}

\author{O. Vince}
\affiliation{Astronomical Observatory, Volgina 7, 11060 Belgrade, Serbia}

\author{M. F. Gu}
\affiliation{Key Laboratory for Research in Galaxies and Cosmology, Shanghai Astronomical Observatory, Chinese Academy
of Sciences, Shanghai 200030, China}

\author{Z. Zhang}
\affiliation{Shanghai Astronomical Observatory, Key Laboratory of Radio Astronomy, Chinese Academy of Sciences, Shanghai 200030, China}

\author{R. Bachev}
\affiliation{Institute of Astronomy and National Astronomical Observatory, Bulgarian Academy of Sciences,
72 Tsarigradsko Shosse Blvd., 1784 Sofia, Bulgaria}

\author{R. Chanishvili}
\affiliation{Abastumani Observatory, Mt. Kanobili, 0301 Abastumani, Georgia}

\author{R. Itoh}
\affiliation{Department of Physics, Tokyo Institute of Technology, 2-12-1 Ookayama, Meguro-ku, Tokyo 152-8551, Japan}

\author{M. Kawabata}
\affiliation{Hiroshima Astrophysical Science Center, Hiroshima University, Kagamiyama 1-3-1, Higashi-Hiroshima
739-8526, Japan}

\author{S. O. Kurtanidze}
\affiliation{Abastumani Observatory, Mt. Kanobili, 0301 Abastumani, Georgia}

\author{T. Nakaoka}
\affiliation{Hiroshima Astrophysical Science Center, Hiroshima University, Kagamiyama 1-3-1, Higashi-Hiroshima
739-8526, Japan}

\author{M. G. Nikolashvili}
\affiliation{Abastumani Observatory, Mt. Kanobili, 0301 Abastumani, Georgia}

\author{{\L}. Stawarz}
\affiliation{Astronomical Observatory of Jagiellonian University, ul. Orla 171, 30-244 Krakow, Poland}

\author{A. Strigachev}
\affiliation{Institute of Astronomy and National Astronomical Observatory, Bulgarian Academy of Sciences,
72 Tsarigradsko Shosse Blvd., 1784 Sofia, Bulgaria}

\begin{abstract}
\noindent
We report on a
recent multi-band optical photometric and polarimetric observational campaign of the blazar 
OJ 287 which was carried out during September 2016 -- December 2017. We employed nine telescopes in Bulgaria, 
China, Georgia, Japan, Serbia, Spain and the United States.  We collected over 1800 photometric image frames 
in BVRI bands and over 100 polarimetric measurements over $\sim$175 nights. In 11 nights with many 
quasi-simultaneous multi-band (V, R, I) observations, we did not detect any genuine intraday variability in flux 
or color. On longer timescales, multiple flaring events were seen. Large changes in color with respect to time 
and in a color--magnitude diagram were seen, and while only a weak systematic variability trend was noticed in color 
with respect to time, the color--magnitude diagram shows a  bluer-when-brighter trend.   
Large changes in the degree of polarization, and substantial swings in the polarization angle were detected. 
The  fractional Stokes parameters of the polarization showed a systematic trend with time in the beginning of 
these observations, followed by chaotic changes and then an apparently systematic variation at the end.
These polarization changes coincide with the detection and duration of the source at very high energies as seen 
by VERITAS. The spectral index shows a systematic variation with time and V-band magnitude. We briefly discuss  
possible physical mechanisms that could explain the observed flux, color, polarization, and spectral variability.
\end{abstract}


\section{Introduction}
\noindent
Blazars comprise a subclass of radio-loud active galactic nuclei in which one of the relativistic jets emanating
from the super massive black hole (SMBH) of mass 10$^{6}$ -- 10$^{10}$ M$_{\odot}$ is pointed close to the observer 
\citep{2002ApJ...579..530W}. This class is composed of BL Lac objects, which have featureless or very weak emission 
lines (equivalent widths, EW $\leq$ 5\AA) \citep{1991ApJS...76..813S,1996MNRAS.281..425M} and flat spectrum radio 
quasars (FSRQs) which have prominent emission lines \citep{1978PhyS...17..265B,1997A&A...327...61G}. 
Blazars show flux variations across the complete 
electromagnetic (EM) spectrum on all possible time scales, i.e. as short as a few minutes to as long as many 
years. They show variable polarization in radio to optical bands, and their emission across the EM spectrum is 
predominantly non-thermal. Their multi-wavelength (MW) spectral energy distribution (SED) is a
double humped structure in which the low energy hump peaks somewhere in IR to soft X-rays and is due to 
synchrotron emission from non-thermal electrons in the jet while  the high energy hump peaks in GeV to TeV 
energies and is probably due to inverse-Compton up-scattering of synchrotron (SSC, synchrotron self Compton) 
or external photons (EC, external Compton) by the relativistic electrons producing the synchrotron emission 
\citep{1998A&A...333..452K,2010ApJ...718..279G}. \\
\\
In the age of multi-wavelength (MW) transient astronomy, blazars are among the best types of persistent,
highly variable, but non-catastrophic, sources for which simultaneous MW observations should be 
performed in order to understand 
their emission mechanism over the complete EM spectrum.  Flux and polarization variations in the range of 
minutes to less than a day are commonly called as intraday variability \citep[IDV;][]{1995ARA&A..33..163W} 
or microvariability \citep{1989Natur.337..627M} or intranight variability \citep{2012A&A...544A..37G}, 
while those with timescales from days to a few months is called short term variability (STV), and timescales 
of several months to years is known as long term variability \citep[LTV;][]{2004A&A...422..505G}. There is 
a lengthy series of papers in which blazars' optical flux and polarization 
variability on diverse timescales are reported 
\citep[e.g.][and references therein]{2003A&A...409..857A,2011A&A...531A..38A,2006A&A...450...39G,2007MNRAS.381L..60C,
2012AJ....143...23G,2012MNRAS.420.3147G,2012MNRAS.425.3002G,2014ApJ...781L...4G,2015MNRAS.452.4263G,
2008AJ....135.1384G,2012MNRAS.425.1357G,2016MNRAS.458.1127G,2017MNRAS.465.4423G,2017MNRAS.472..788G,
2016MNRAS.461.3047L,2018MNRAS.473.1145K}. \\ 
\\
The blazar OJ 287 ($z=0.306$), though identified in 1967 \citep{1967AJ.....72..757D}
has had data taken in the optical bands since $\sim$1890, and by using about a
century long light curve (LC), \citep{1988ApJ...325..628S} noticed that it showed
double peaked outbursts almost every 12 years. To explain them, they proposed a
binary black hole model and predicted the next outbursts would occur in 1994--1995.
An extensive optical monitoring campaign known as OJ-94 was organized around the
globe and the predicted double peaked outbursts were indeed detected in 1994--1995, 
separated by $\sim$ 1.2 years \citep[e.g.][]{1996A&A...305L..17S,1996A&A...315L..13S}. 
In the next intense observing campaign on OJ 287 during 2005 -- 2007, the double peaked outbursts were
again detected, with the first one at the end of 2005 and the
second  at the end of 2007, separated by $\sim$2 years \citep{2009ApJ...698..781V}.
For the most recent prediction of double peaked outbursts, the first
outburst was detected in December 2015 while the second outburst still has
to be detected \citep{2016ApJ...819L..37V,2017MNRAS.465.4423G} 
and is predicted for mid-2019 \citep{2016ApJ...819L..37V}. \\
\\
The most puzzling issues in the double peaked outbursts of OJ 287 are the timing of the
detection of the second outburst and its strength. From the last three sets of outbursts detected
since 1994, it is now clear that they are not exactly periodic. \citet{1996ApJ...460..207L}  
analyzed the substructure of major outbursts of OJ 287,
identified sharp flares and connected these with a model in which a secondary SMBH crosses
the accretion disk of the primary SMBH during their mutual binary orbit. They estimated the 
masses of the primary and secondary SMBHs to be 17 $\times$
10$^{9}$ M$_{\odot}$ and 10$^{8}$ M$_{\odot}$, respectively. The original model of 
\citet{1988ApJ...325..628S}
has been modified in different ways over the past decade. \citet{2008Natur.452..851V} 
claimed that the changing binary system provides evidence for the loss of orbital energy to within
$\sim$10\% of the value predicted by the quadrupole formula for the emission of gravitational waves from the system.  
However, using 2015 data and considering the higher order radiation terms, a more recent analysis claimed the loss of orbital energy 
to be $\sim$6\% less than the quadrupole formula indicates \citep{2018ApJ...866...11D}.  
\citet{2008Natur.452..851V,2010ApJ...709..725V} 
explain the deviations of the outbursts from strict periodicity as arising from this gravitational wave
driven in-spiraling of the  binary black hole system present at the centre of OJ 287.
According to the latest iteration of the model, the source is an inspiraling and
precessing binary SMBH system with a current period of 12.055 years which decreases
by 38 days/century and involves a minimum separation of 1.1 years between the twin outbursts
associated with the disk impacts \citep{2017Galax...5...83V}. \\
\\
The recent high activity phase of OJ 287 started in November 2015, around the anticipated
time of the latest predicted disk impact optical outburst \citep{2016ApJ...819L..37V,
2017MNRAS.465.4423G}. The outburst was detected on December 2015 and was the brightest
in the last three decades with a relatively low polarization fraction (PD) of $<10\%$
\citep{2016ApJ...819L..37V,2017MNRAS.465.4423G,2018MNRAS.473.1145K}, as was expected
in the binary SMBH model. However, the other key polarization property, the polarization
angle (PA), showed an extraordinary $\rm\sim~200^o$ systematic change over the duration of
the optical flare \citep{2018MNRAS.473.1145K}. Similar characteristics (a low PD,
but strong change in PA), were also seen during the first flare of the 1993--1994 outbursts
\citep{2000A&AS..146..141P}, while there were not enough observations during the 2005
outburst to know if this also was the case then \citep{2010MNRAS.402.2087V}. This relatively lower polarization compared
to that of other flares following it, has been argued to be a clear signature of thermal
emission \citep{2016ApJ...819L..37V}. Additionally, a multi-wavelength investigation
of the NIR-optical SEDs during this duration made in our previous work \citep{2018MNRAS.473.1145K},
 has, for the first time, reported a bump in the NIR-optical region, consistent with
the standard accretion disk impact description of the primary SMBH. Exploration of optical data by
\citet{2017MNRAS.465.4423G}  found many nights showing IDV during this period.
\\
Using data from the last three
outbursts detected since 1994, the masses of the primary and secondary black holes
have been estimated to be (1.83$\pm$0.01) $\times$ 10$^{10}$ M$_{\odot}$ and (1.5$\pm$0.1)
$\times$ 10$^{8}$ M$_{\odot}$, respectively, and the spin of primary black hole was
claimed to be 0.313$\pm$0.01 \citep{2016ApJ...819L..37V}. A strong flare detected
in March 2016 was comparably strong to the December 2015 outburst and had a similar 
polarization \citep{2018MNRAS.473.1145K,2017MNRAS.465.4423G}. In general, blazars,
including OJ 287, can evince large amplitude flares with wide ranges in polarization properties,
so determining which of them are those outbursts that are actually caused by the
impacts associated with the binary black hole model remains a problem. The 
binary SMBH model would expect that those particular outbursts
are distinguished by a comparatively low optical polarization \citep{2008A&A...477..407V}. \\ 
\\
In the present work, we report detailed optical flux and polarization measurement taken during September
2016 -- December 2017 of the blazar OJ 287. This is a continuation of an ongoing optical monitoring campaign
around the globe of OJ 287 since the year 2015. In that earlier work we detected several flares in flux 
and significant changes in the degree of polarization and polarization 
angle \citep{2017MNRAS.465.4423G,2018MNRAS.473.1145K}. We will continue our observing campaign on 
this blazar, at least until we see if we detect the second predicted  outburst of the current pair in 2019. \\
\\
We structured the paper as follows. In Section 2, we provide information about our new optical photometric and 
polarimetric observational data and its analysis. In Section 3, we present the results, and we discuss them in
Section 4. We summarize our results in Section 5. \\    
\\
\section{Observations and Data Reductions}

\noindent
Our new optical photometric and polarimetric observational campaigns were carried out from September 2016
to December 2017. Photometric observations were carried out using seven telescopes located in 
Japan, China, Bulgaria (2 telescopes), Georgia, Serbia, and Spain. Using these seven telescopes, 
photometric observations were taken on 174 observing nights during which we collected a total of 1829 
image frames of OJ 287 in B, V, R, and I optical photometric bands. \\
\\
We also used the archival optical photometric and polarimetric observations which are performed at Steward 
Observatory, University of Arizona, using the 2.3-m Bok and 1.54-m Kupier telescopes. Polarimetric observations 
were carried out over 94 observing nights for which there were a total of 104 polarimetric measurements of OJ 287.
We call these two telescopes collectively as telescope A in the photometric observation log provided in Table 1. 
The observations from them are taken from the public 
archive\footnote{http://james.as.arizona.edu/$\sim$psmith/Fermi/datause.html}. 
These photometric and polarimetric observations of OJ 287 were carried out using SPOL CCD Imaging/Spectropolarimeter 
attached to those two telescopes. Details about the instrument, observation and data analysis are provided in 
detail in \citet{2009arXiv0912.3621S}. \\ 
\\
Observations from Weihai observatory of Shandong University  employed the 1.0-m telescope at Weihai, China. 
This telescope is named as Telescope B in the observation log provided in Table 1. It is a classical 
Cassegrain telescope with a focal ratio of f/8. The telescope is equipped with a back-illuminated Andor 
DZ936 CCD camera and BVRI filters. We provide critical information about the telescope and CCD detector 
in Table 2 and additional details are given in \citet{2014RAA....14..719H}. 
Sky flats for each filter were taken at twilight, and usually 10 bias frames were taken at the beginning 
and the end of the observation. All frames were processed automatically by using an Interactive Data Language 
(IDL) procedure developed locally which is based on the NASA IDL astronomical 
libraries\footnote{http://idlastro.gsfc.nasa.gov/}. 
Firstly, all frames were bias and flat-field corrected. Secondly, the magnitude was derived by differential 
photometry technique using local standard stars 4, 10 and 11 in the blazar field \citep{1996A&AS..116..403F}. 
The photometry radius was set to 14 pixels, and the inner and outer radii for sky brightness were set to 
30 and 40 pixels, respectively. Most of our intensive observations targeting IDV were done using this
telescope. \\

\begin{table}
\noindent
{\bf Table 1.} Observation log of optical photometric observations of the blazar OJ 287. \\
\\
\centering
\begin{tabular}{lcclcclcclcc} \hline
~~~~Date & Telescope  & Data Points \\
yyyy mm dd  &     &B, V, R, I \\
\hline
2016 09 24  &  A  &0, 1, 1, 0 \\
2016 09 25  &  A  &0, 1, 1, 0 \\
2016 10 09  &  B  &0, 1, 1, 1 \\
2016 10 10  &  B  &0, 1, 1, 1 \\
\hline
\end{tabular} 

\noindent
(This table is available in its entirety in a machine-readable form in the online journal. A portion is
shown here for guidance regarding its form and content)
\end{table}

\begin{table*}
{\bf Table 2.} Details of new telescopes and instruments
\hspace*{-0.2in}
\centering
\begin{tabular}{llll} \hline
                & Weihai, China                   & Las Casqueras, Spain              &  ASV, Serbia \\\hline
Telescope       & 1.0-m Cassegrain                & 35.6 cm Schmidt Cassegrain        &  1.4-m RC Nasmyth \\
CCD Model       & Andor DZ936                     & ATIK 383L+ Monochrome             &  Apogee Alta U42 \\
Chip Size       & 2048 $\times$ 2048 pixels$^{2}$ & 3354 $\times$ 2529 pixels$^{2}$   &  2048 $\times$ 2048 pixels$^{2}$ \\
Scale           & 0.35 arc sec pixel$^{-1}$       & 1.38 arc sec pixel$^{-1}$         &  0.243 arc sec pixel$^{-1}$ \\
Field           & 12 $\times$ 12 arcmin$^{2}$     & 25.46 $\times$ 19.16 arcmin$^{2}$ &  8.3 $\times$ 8.3 arcmin$^{2}$ \\
Gain            & 1.8 e$^{-1}$ ADU$^{-1}$         & 0.41 e$^{-1}$ ADU$^{-1}$          &  1.25 e$^{-1}$ ADU$^{-1}$ \\
Read out noise  & 7 e$^{-1}$ rms                  & 7 e$^{-1}$ rms                    &  7 e$^{-1}$ rms    \\
Typical Seeing  & 1.3 -- 2 arcsec                 & 1.5 -- 2.5 arcsec                 &  1 -- 1.5 arcsec \\\hline
\end{tabular}
\end{table*}

\noindent
Our optical photometric observing campaign of the blazar OJ 287 in the B, V, R, and I passbands continued to
use two telescopes in Bulgaria (2.0 m and 50/70 cm Schmidt) which are conflated as Telescope C in Table 1. 
These telescopes are equipped with CCD detectors and broad-band optical filters B, V, R, and I. Details of 
these telescopes and the CCDs mounted on telescopes C as well as details of the reduction procedures used 
are given in our earlier papers \citep{2015MNRAS.451.3882A,2016MNRAS.458.1127G}. \\ 
\\
Optical V band photometric observations of the blazar OJ 287 were carried out using a Celestron C14 XLT 35.6cm 
with reducer f/6.3 located at Las Casqueras, Spain. The telescope is equipped with CCD camera and V broad band 
optical filter and is called Telescope D in Table 1. Details about this telescope and CCD are given in Table 2. 
Standard image processing (bias, flat field and dark corrections) are applied and the photometry data were reduced 
using the Software MaxIm DL. Reference stars  available in the database of the American Association of Variable 
Star Observers (AAVSO)\footnote{https://www.aavso.org/apps/vsp/} are used for calibrating the V magnitude of OJ 287. \\ 
\\
For this campaign, observations of OJ 287 were also carried out using the newly installed 1.4-m telescope at  
Astronomical Station Vidojevica of the Astronomical Observatory in Belgrade (ASV). The telescope is equipped with CCD 
camera and B, V, R, I broad band optical Johnson-Cousins filters. Details are given in Table 2. During our 
observations, the CCD was cooled to 30$^{\circ}$C below ambient. The camera is back-illuminated with high 
quantum efficiency (QE):peak QE at 550 nm $>$ 90\%. The observations carried out by this  telescope are given 
in the observation log reported in Table 1 where it is denoted as Telescope E. 
Photometric observations were done in 1 $\times$ 1 binning mode in B, V, R, I passbands. Standard optical
photometric data analysis procedures were adopted (e.g.\ bias and flat field correction). To obtain instrumental 
B, V, R, I magnitudes of OJ 287 and comparison stars, usually we took 3 image frames per filter, and the result 
is the average value of the estimated magnitude. Local standard stars in the blazar OJ 287 field are used 
to calibrate the magnitude of OJ 287 \citep{1996A&AS..116..403F}. \\
\\
Some data from the 70 cm telescope at Abastumani Observatory in Georgia were taken in R band and they are listed 
as Telescope F in the photometric observation log given in Table 1. The  1.5 m telescope in Kanata, Japan is named 
as Telescope G in that table and it also contributed data in V and R bands on a few nights. Details about telescopes 
F and G, their CCDs, broad band filters, and the data reduction employed are given in our earlier paper 
\citep{2017MNRAS.465.4423G}. \\

\section{Results}

\begin{figure*}
\centering
\includegraphics[scale=0.8]{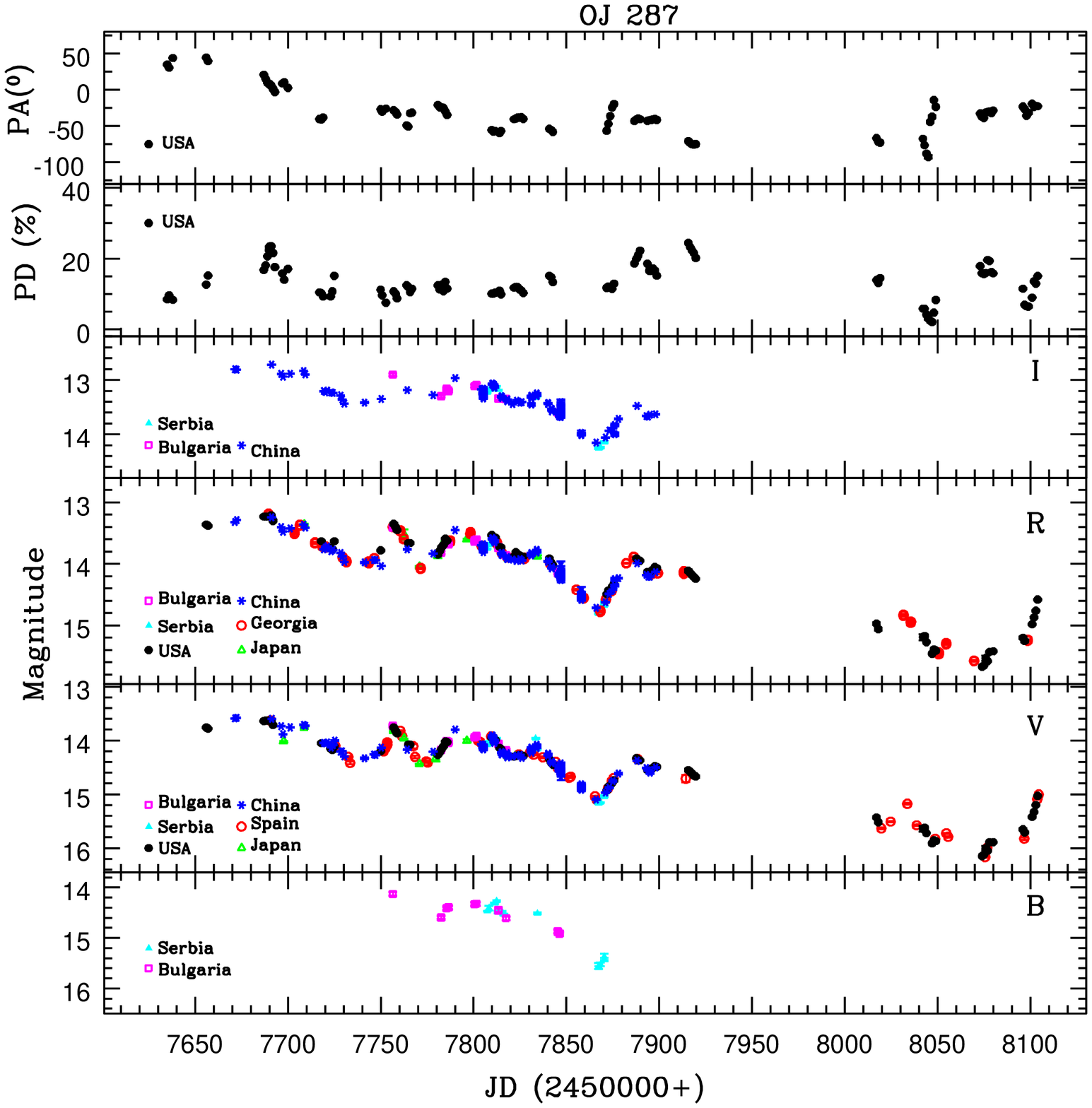}
\caption{Optical flux and polarization variability light curves of OJ 287 during September 2015 -- December 2017.
From bottom to top, the panels  show B, V, R and I calibrated magnitudes, 
degree of polarization in R band, and polarization angle in R band, respectively. 
Different symbols and colors marked inside the panels indicate the data from different telescopes.}
\end{figure*}

\noindent
In Fig.\ 1 we present  the photometric and polarimetric light curves (LCs) generated from
our observing campaign using nine telescopes around the globe during September 2016 -- December 2017. 
We present  LCs of the B, V, R, and I bands, as well as the degree of polarization (PD) and
polarization angle (PA). The V and R band LCs clearly have the densest observational cadence.  
The polarimetric observations do not have similarly dense coverage though the most common photometric
observations are made in the R band. One can immediately notice that there are several flaring events 
in the photometric observations in V and R bands, most of which are also seen in I, as well as large changes in the degree of polarization 
and polarization angle. In the
following subsections, we discuss the variability characteristics of the blazar OJ 287 on IDV, STV and
LTV timescales, and the nature of the polarization variation. 

\subsection{Light Curve Analysis Techniques}

\noindent
To quantify the IDV variability results, we use variability detection techniques based on the 
{\it F}-test, the so-called $\chi^{2}$-test, and the Levene test and we describe them briefly in the following 
subsections. The {\it F}-test and $\chi^{2}$-test, which are frerquently employed in studies of AGN variability, assume a normal distribution of the data which is not, 
in general, true for blazars' LCs. Thus, we have additionally performed the Levene test, which is a nonparametric variance 
test. 
We conservatively claim a LC is variable only if the variability is detected by all three tests. We also 
calculate the variability amplitude on IDV, STV and LTV timescales. The method used to determine the 
variability amplitude is also described  briefly below. 
 
\subsubsection{F-Test}

\noindent
To quantify any IDV of the blazar OJ 287, we originally adopted the commonly used {\it F}-Test \citep{2010AJ....139.1269D} and display these results because many other papers in this field have done so, even though it relies on the data displaying a normal distribution.
We employ two comparison stars and so define 
\citep[][and the references therein]{2012AJ....143...23G,2015MNRAS.450..541A,2017MNRAS.465.4423G},

\begin{equation}
 \label{eq.ftest}
 F_1=\frac{Var(BL - star 1)}{Var(star 1 - star 2)}, 
 F_2=\frac{Var(BL - star 2)}{Var(star 1 - star 2)}
\end{equation}
where (BL $-$ star 1), (BL $-$ star 2), and (star 1 $-$ star 2) are the differential instrumental magnitudes
of blazar and standard star 1, blazar and standard star 2, and standard star 1 and standard star 2, respectively, 
while Var(BL $-$ star 1), Var(BL $-$ star 2), and Var(star 1 $-$ star 2) are the variances of the differential 
instrumental magnitudes. \\
\\
Then the relevant $F$ value is the average of $F_1$ and $F_2$ which is then compared with the critical $F^{(\alpha)}_{\nu_{bl},\nu_*}$
value where $\alpha$ is the significance level set for the test while $\nu_{bl}$ and $\nu_*$ are the number of
degrees of freedom, calculated as ($N - 1$), with $N$  the number of measurements. For IDV detection in the LCs, we have 
done the $F$-test for $\alpha$ values of 0.999 and 0.99, which effectively correspond to $3 \sigma$ and $2.6 \sigma$ 
detections, respectively. 
The null hypothesis (no variability) is discarded if the $F$ value 
is greater than the critical value, and as usual we claim a LC to 
be variable if $F > F_c(0.99)$ \citep{2015MNRAS.450..541A,2017MNRAS.465.4423G}. 

\begin{figure*}
\centering
\epsfig{figure= 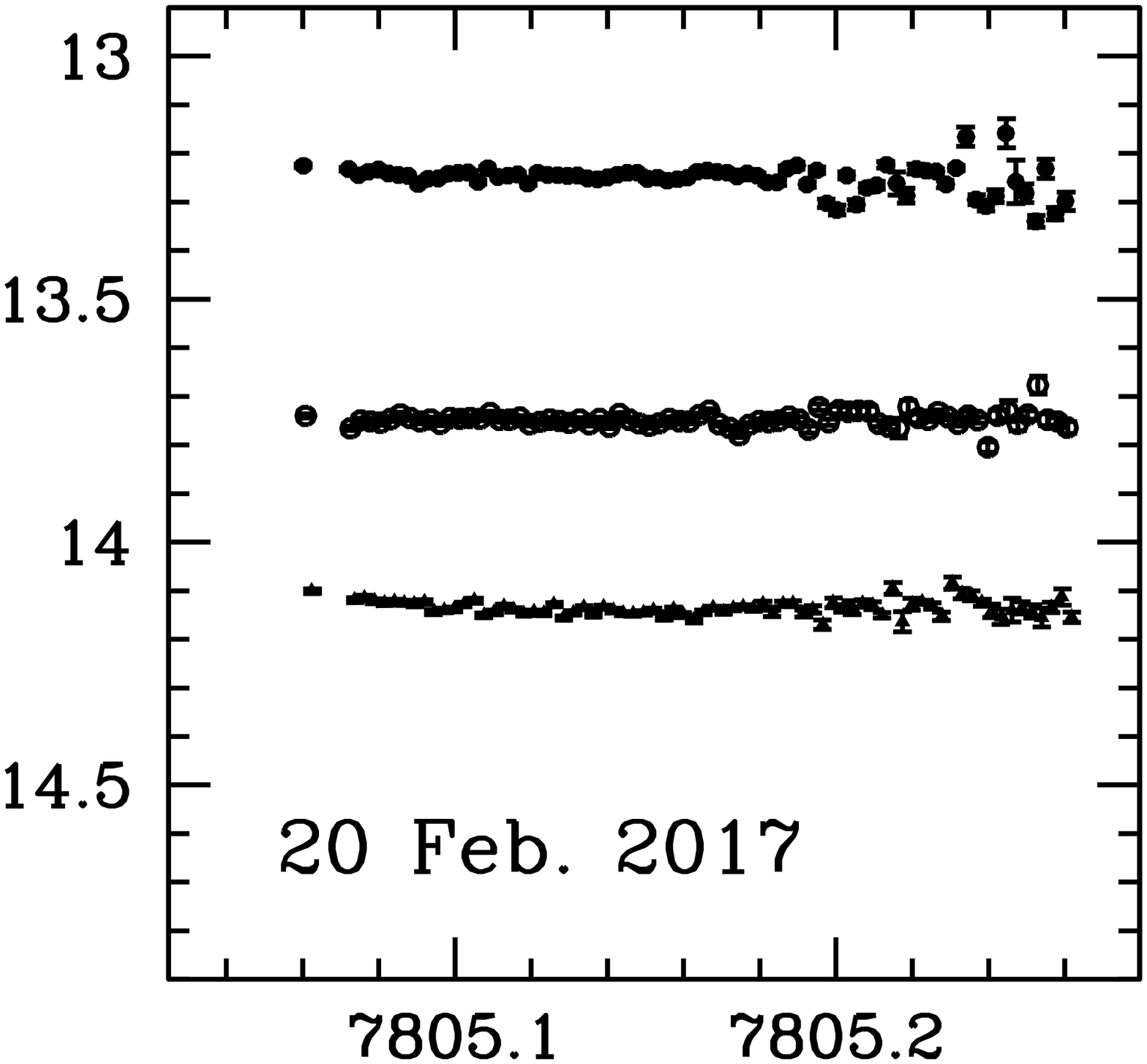,height=2.0in,width=2.0in,angle=0}
\epsfig{figure= 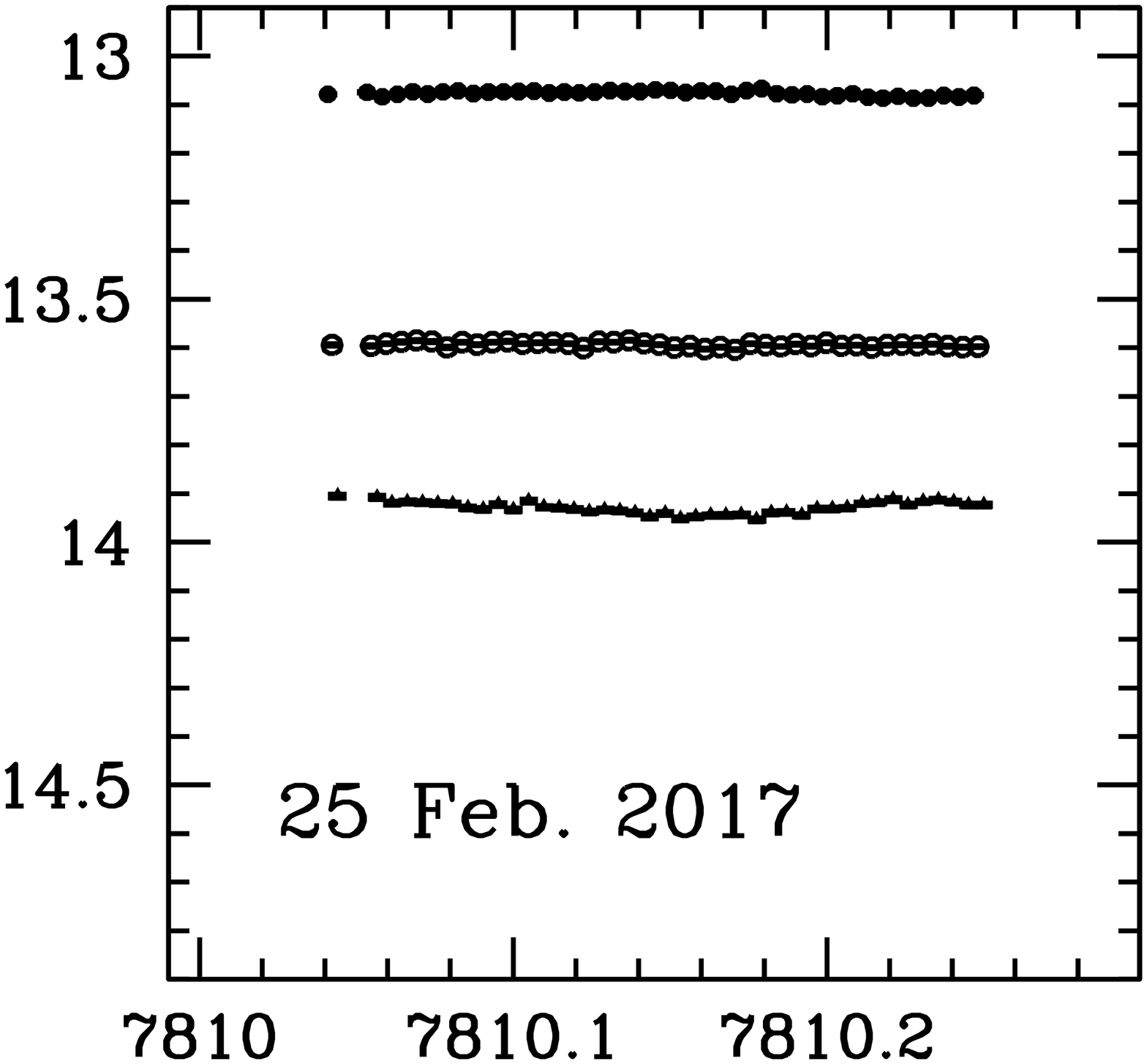,height=2.0in,width=2.0in,angle=0}
\epsfig{figure= 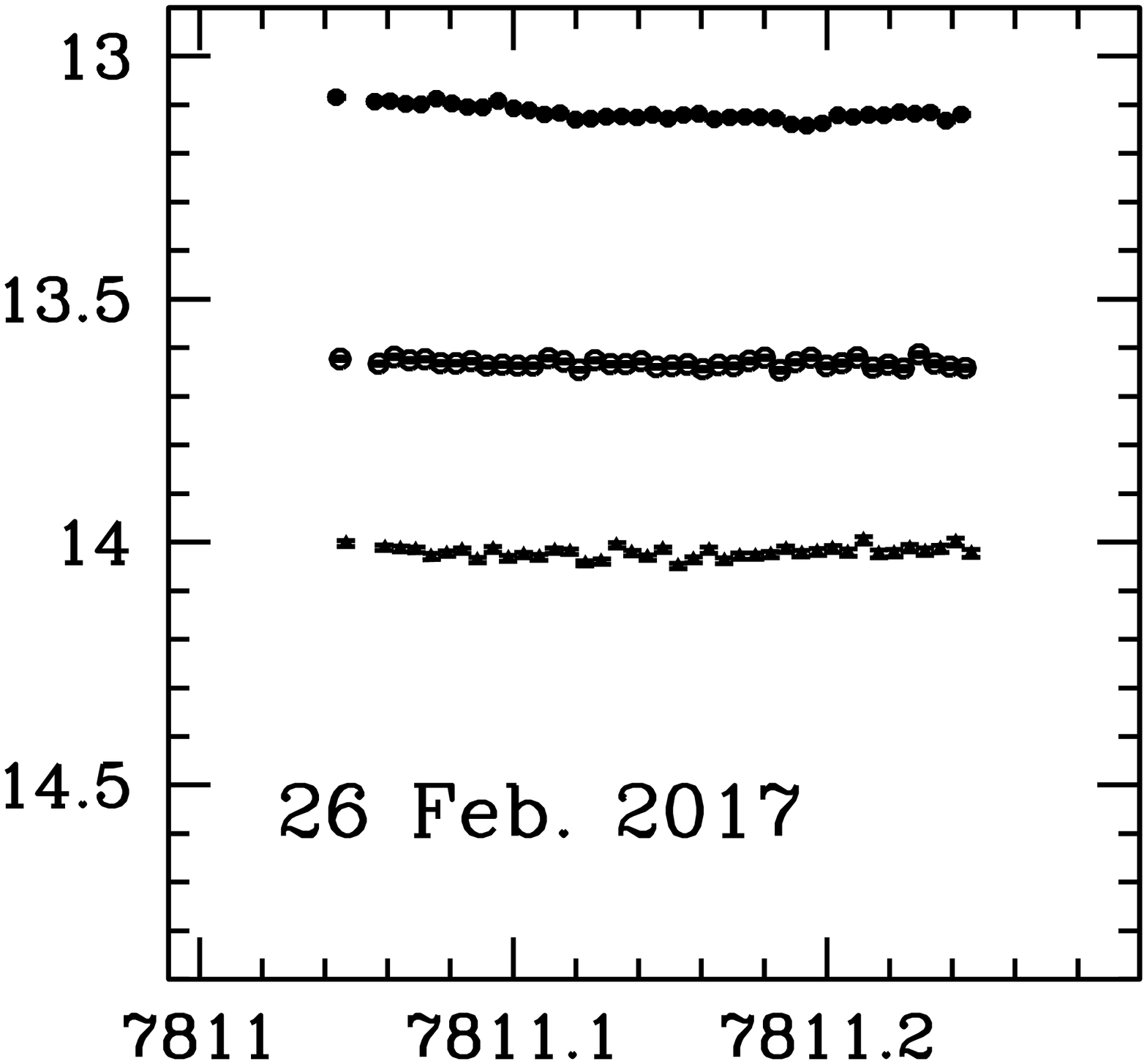,height=2.0in,width=2.0in,angle=0}
\epsfig{figure= 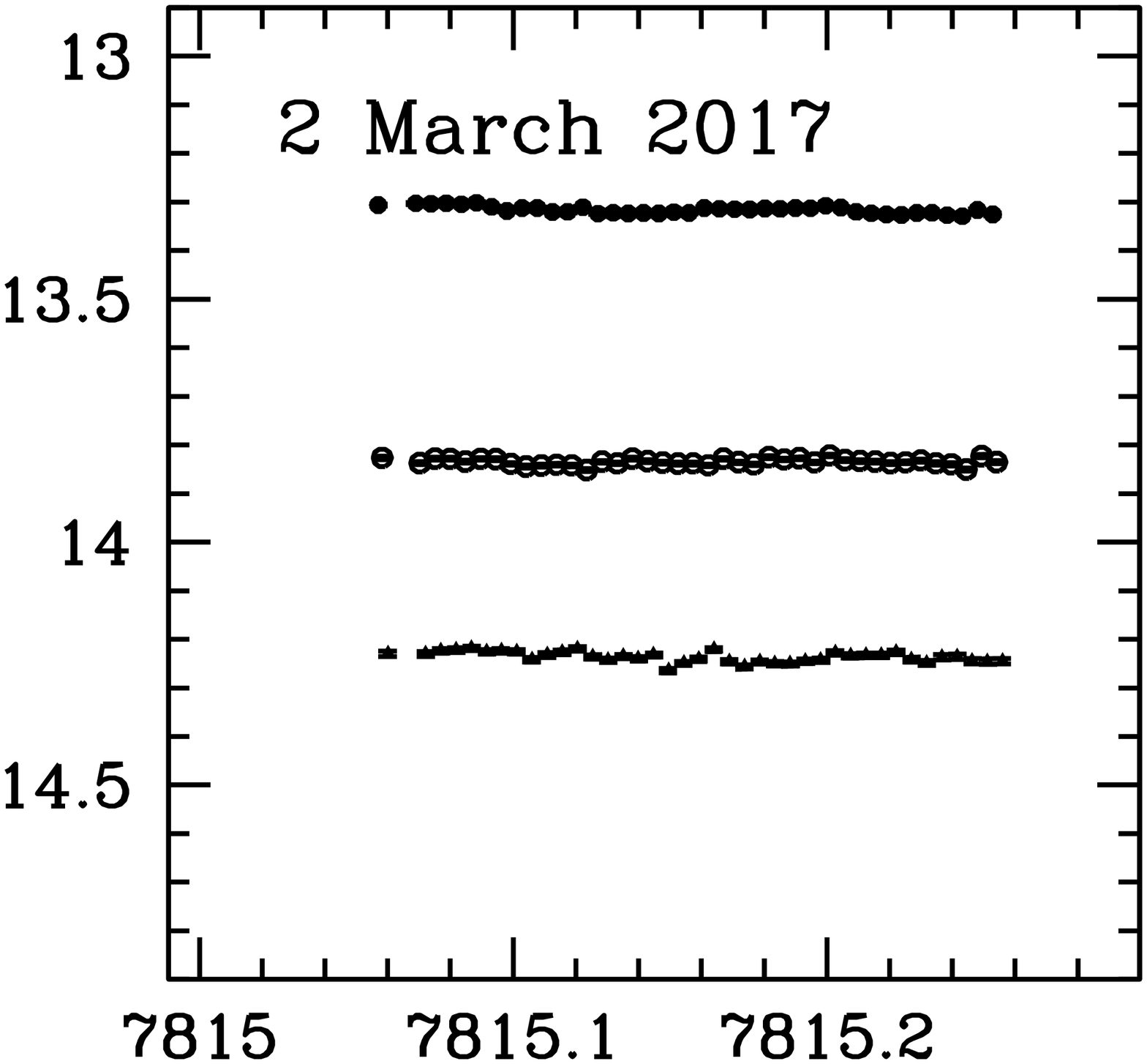,height=2.0in,width=2.0in,angle=0}
\epsfig{figure= 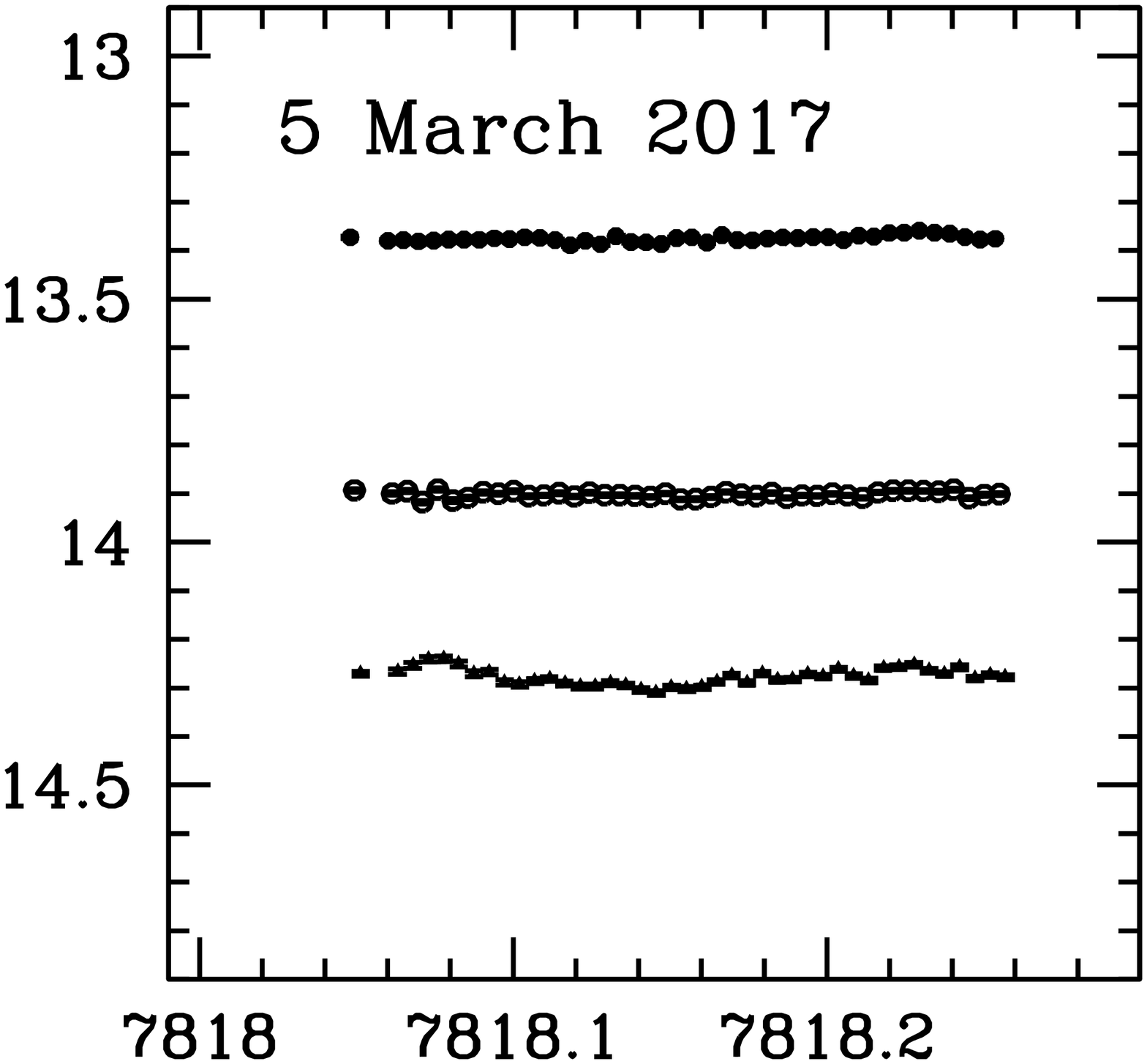,height=2.0in,width=2.0in,angle=0}
\epsfig{figure= 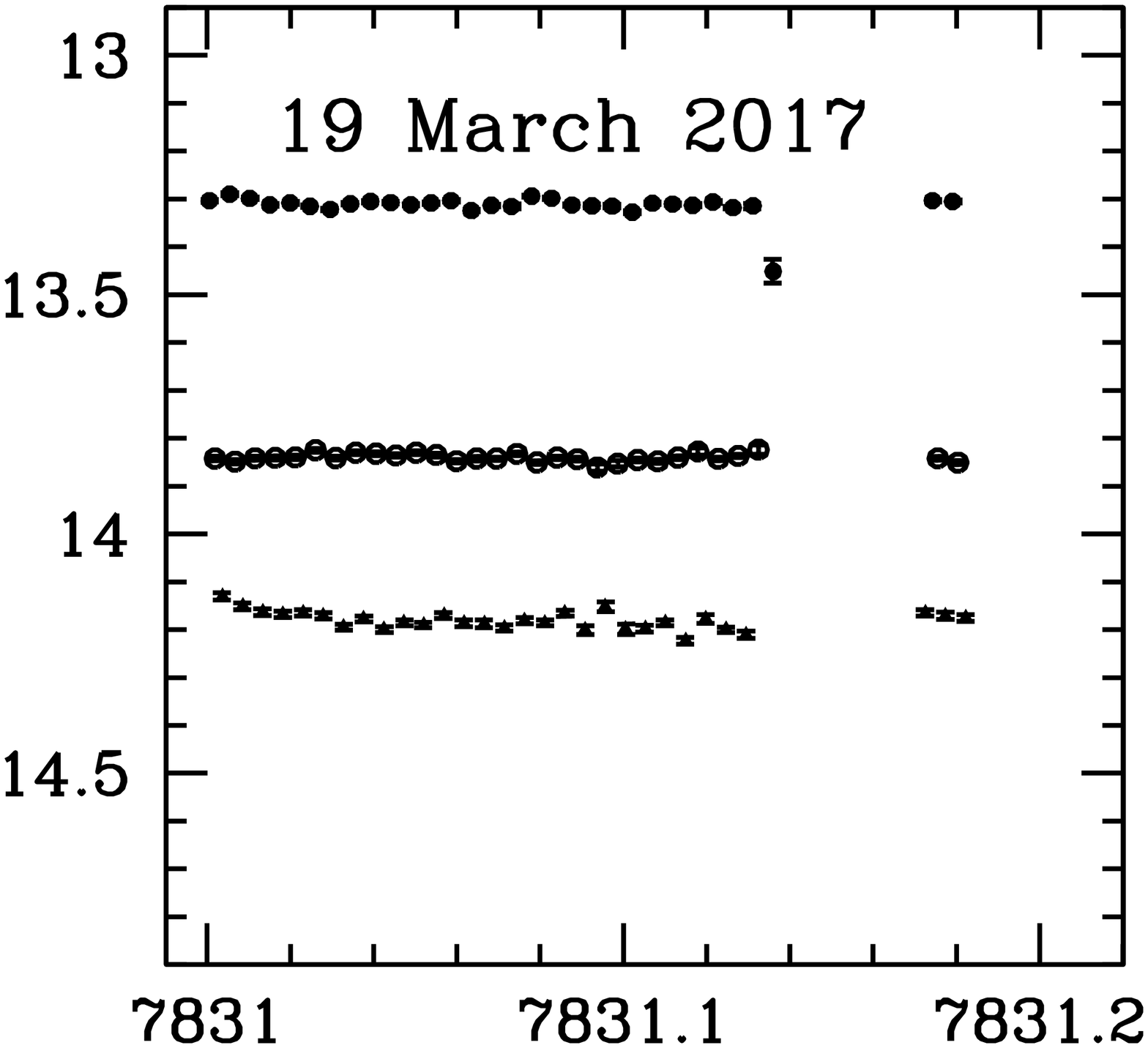,height=2.0in,width=2.0in,angle=0}
\epsfig{figure= 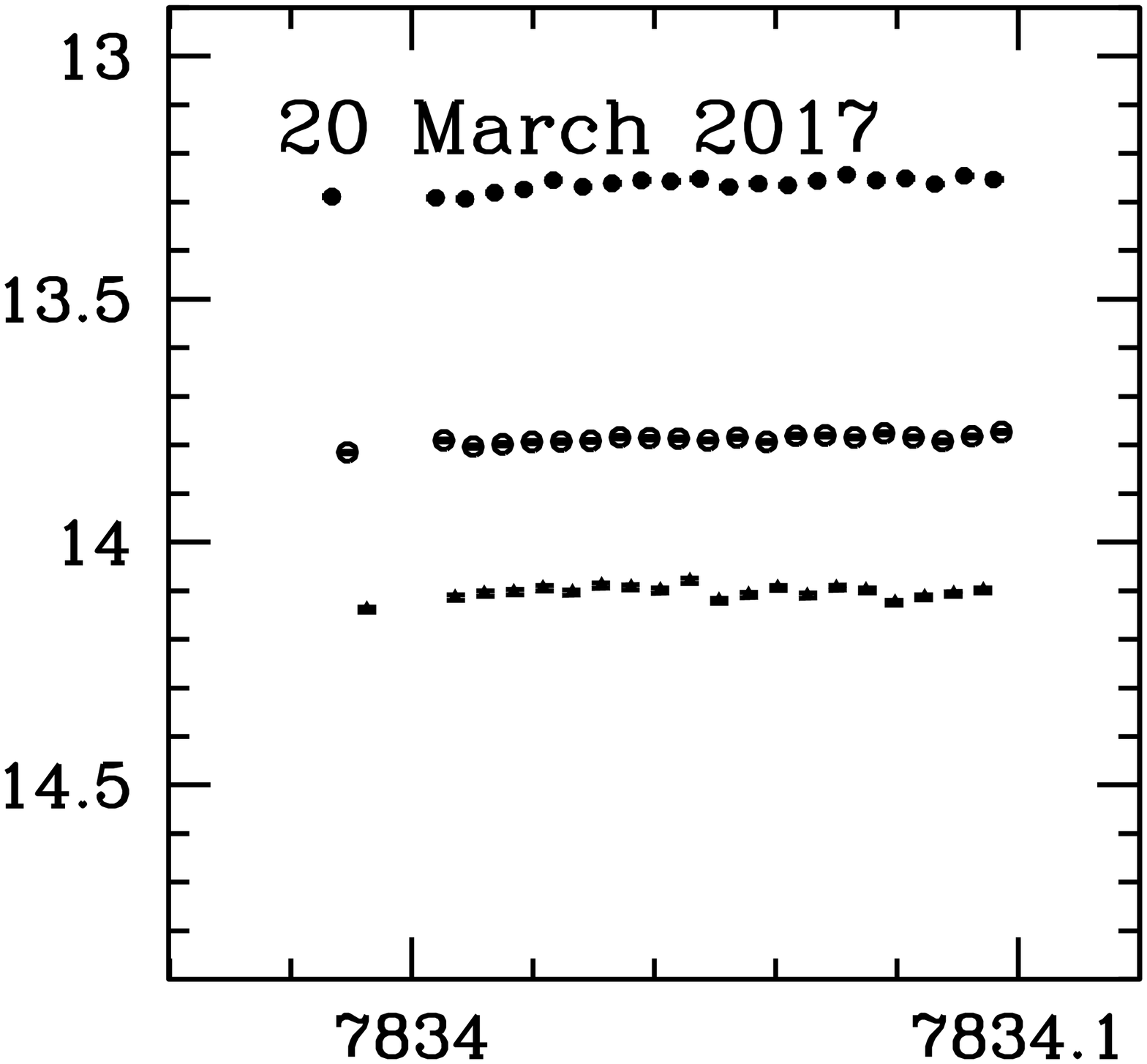,height=2.0in,width=2.0in,angle=0}
\epsfig{figure= 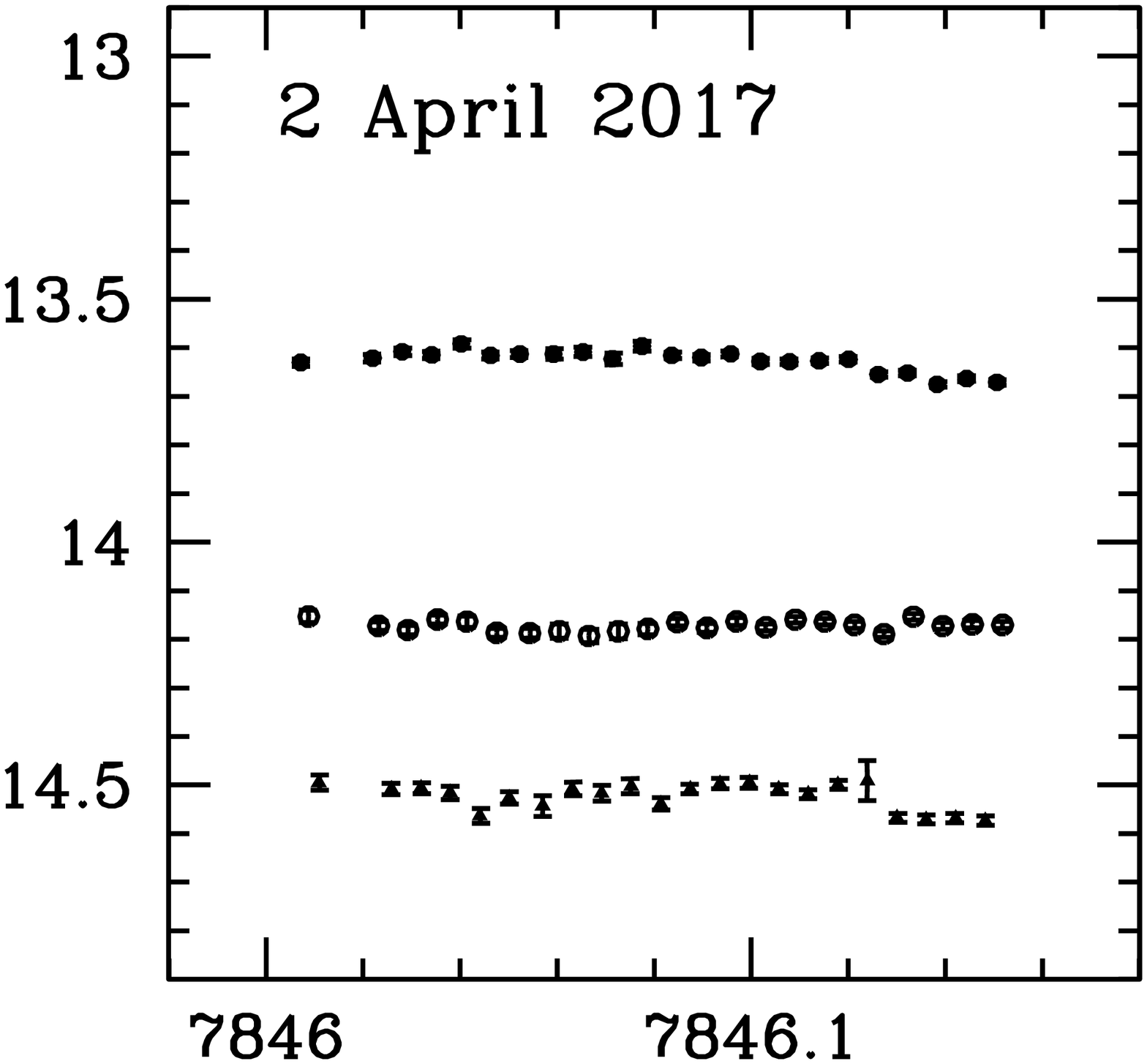,height=2.0in,width=2.0in,angle=0}
\epsfig{figure= 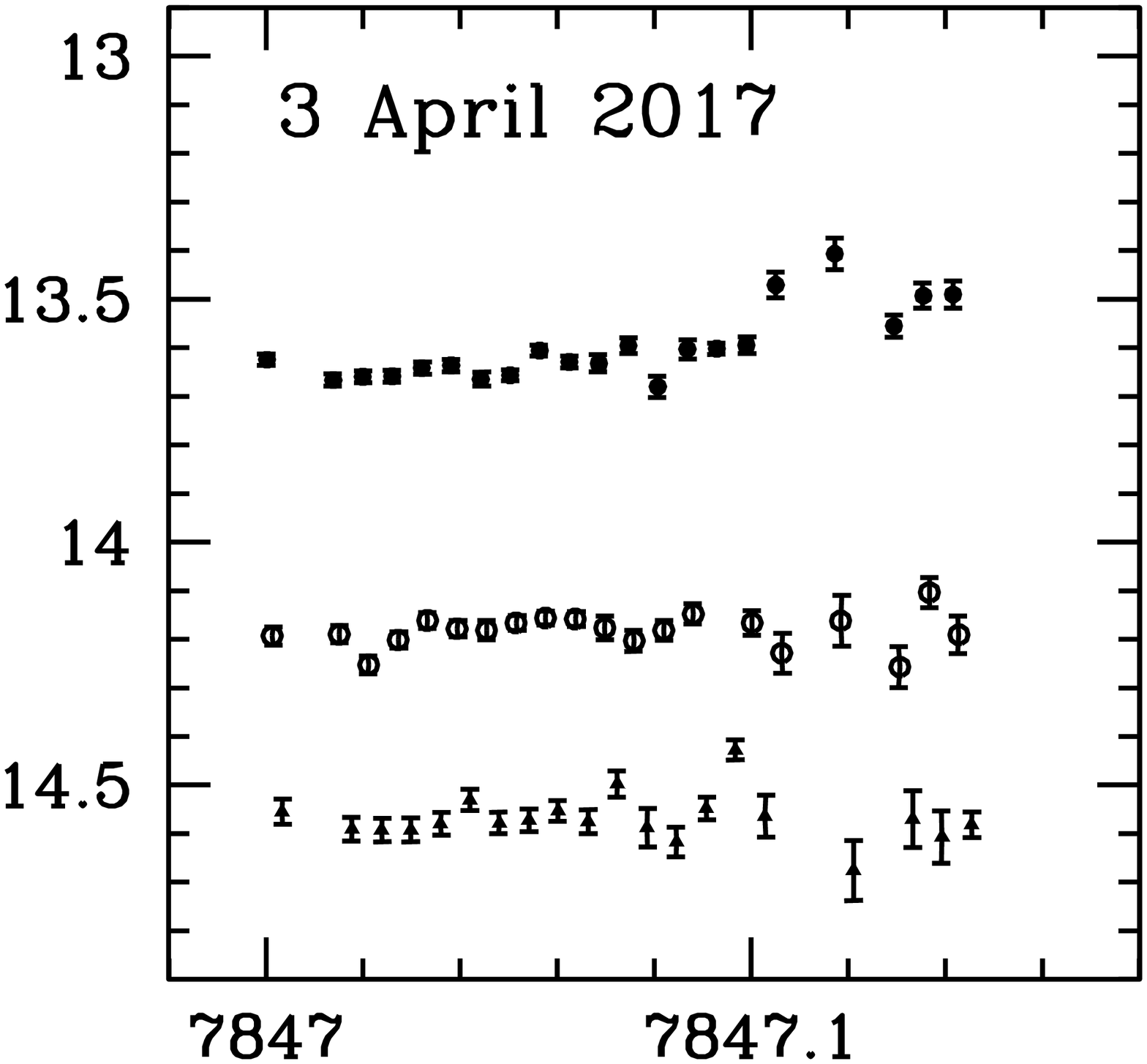,height=2.0in,width=2.0in,angle=0}
  \caption{Intraday light curves for OJ 287 in V, R, I filters. Filled triangles, open circles and filled circles represent data
in V, R, and I filters, respectively. The X axis is JD (2450000+) and the Y axis is the magnitude in each plot, where observation dates are 
indicated in each plot.}
\end{figure*}

\subsubsection{$\chi^{2}$-Test}

\noindent
To quantify the detection of variability of the blazar we have also used the ``so-called'' $\chi^{2}$-test \citep{2010AJ....139.1269D}. 
This $\chi^{2}$ statistic is defined as \citep[e.g,][]{2015MNRAS.450..541A} 

\begin{equation}
\chi^2 = \sum_{i=1}^N \frac{(V_i - \overline{V})^2}{\sigma_i^2},
\end{equation}
where, $\overline{V}$ is the average magnitude, and $V_i$ is the magnitude of $i^{th}$ observation with
a corresponding standard error $\sigma_i$.  We note that this test assumes a Gaussian scatter and constant mean, neither of which are generally seen in blazar LCs, and so these results are included only because this approach has often been
used in other studies. It has been determined that the actual measurement errors
are larger than the errors indicated by photometry software by a factor of 1.3 $-$ 1.75 
\citep[e.g,][]{2003ApJ...586L..25G}. So we multiply the errors obtained from the data reductions by a
factor of 1.5 \citep{2004JApA...25....1S} to get better estimates of the real photometric errors. The 
mean value of $\chi^{2}$ is then compared with the critical value $\chi_{\alpha,\nu}^2$ where $\alpha$ is 
the significance level and $\nu = N - 1$ is the number of degrees of freedom. A value
$\chi^2 > \chi_{\alpha,\nu}^2$ implies the presence of variability. \\  

\subsubsection{Levene Test} 

\noindent
To quantify the IDV in the LCs of the blazar OJ 287, we also have used the non-parametric Levene test 
(Brown \& Forsythe 1974). 
It compares the variances of different samples and tests the null hypothesis that all the samples are 
from populations having equal variances. We calculated the statistics $W_1$ and the null hypothesis 
probability $p_1$ for the differential LCs (DLCs) of blazar $-$ Star A and the DLCs of Star A $-$ Star B. 
Similarly we found $W_2$ and $p_2$ for the DLCs of blazar $-$ Star B and that of Star A $-$ Star B. 
A $p$-value greater than 0.01 indicates that the blazar is non-variable with respect to the star.
The test statistic, $W$, is defined as\footnote{https://en.wikipedia.org/wiki/Levene's\_test} 

\begin{equation}
W = {\frac {(N - k)} {(k - 1)}} . {\frac {\sum_{i = 1}^{k} N_{i} {(Z_{i.} - Z_{..})}^{2}} {\sum_{i = 1}^{k} \sum_{j = 1}^{N_i} {(Z_{ij} - Z_{i.})}^{2}}}
\end{equation}

\noindent
where \\
$k$  is the number of different groups to which the sampled cases belong, \\
$N_{i}$ is the number of cases in the i$^{th}$ group, \\
$N$ is the total number of cases in all groups, \\ 
$Y_{ij}$ is the value of the measured variable for the j$^{th}$ case from the i$^{th}$ group, \\
$Z_{ij} = |Y_{ij} - {\bar{Y}_{i.}}|$, ~~~~ ${\bar{Y}_{i.}}|$ is a mean of the i$^{th}$ group, 
$|Y_{ij} - {\tilde{Y}_{i.}}|$, ~~~~ ${\tilde{Y}_{i.}}|$ is a median of the i$^{th}$ group, \\
$Z_{i.} = {\frac {1} {N_{i}}} {\sum_{j = 1}^{N_i}} Z_{ij}$ is the mean of the $Z_{ij}$ for group $i$,  \\
$Z_{..} = {\frac {1} {N}} {\sum_{i = 1}^{k} \sum_{j = 1}^{N_i}} Z_{ij}$ is the mean of all $Z_{ij}$. \\
\\
We claimed that the blazar to be non-variable if it is non-variable with respect to both the comparison 
stars. The values of test statistics and the null hypothesis probabilities are given in Table \ref{tab:IDVtab}.
We list a source as certainly variable (Var) only if it satisfies the criteria of all the three
tests, i.e. $F$-test, $\chi_2$-test, and Levene test. 

\subsubsection{Amplitude of Variability}

\noindent
The percentage of magnitude and color variations on IDV through LTV time scales can be calculated by
using the variability amplitude parameter $A$, which was introduced by \citet{1996A&A...305...42H} 
and defined as

\begin{eqnarray}
A = 100\times \sqrt{{(A_{max}-A_{min}})^2 - 2\sigma^2}(\%) 
\end{eqnarray}

\noindent
Here $A_{max}$ and $A_{min}$ are the maximum and minimum values in the calibrated magnitude and
color of LCs of the blazar, and $\sigma$ is the average measurement error.

\subsection{Intra-day Flux and Color Variability}

\begin{table*}
{\bf Table 3.}Results of IDV Observations. In the Variable column, Var, and NV represent variable, and non-variable, respectively.\\
\vspace*{0.1in}
\noindent
\scriptsize
\centering
{\begin{tabular}{cccccccc}\hline
Date  &  Band   &  $N$  &                 $F-$test                      & $\chi^2-$test           &  \multicolumn{2}{c}{Levene test} &  Variable  \\  \cmidrule[0.03cm](r){6-7}
yyyymmdd &      &       & $F_1$, $F_2$, $F$, $F_c$(0.99), $F_c$(0.999) & $\chi^2_1$, $\chi^2_2$, $\chi^2_{av}$, $\chi^2_{0.99}$, $\chi^2_{0.999}$&  $W_1$,  $p_1$  &  $W_2$, $p_2$ &          \\\hline
20170220 & V &  74 &   0.98,   0.67,   0.83,   1.73,   2.08 &  75.73, 127.15, 101.44, 104.01, 116.09     & 4.66e-01,  4.96e-01  &  6.56e-05, 9.94e-01 & NV \\
         & R &  74 &   1.50,   0.90,   1.20,   1.73,   2.08 &  59.06,  70.95,  65.00, 104.01, 116.09     & 5.29e-01,  4.68e-01  &  2.88e-01, 5.92e-01 & NV \\
         & I &  74 &   1.18,   0.22,   0.70,   1.73,   2.08 & 320.75, 102.60, 211.67, 104.01, 116.09     & 5.56e-03,  9.41e-01  &  1.99e+01, 1.59e-05 & NV \\
         &(V-R) &  74 &   1.00,   0.61,   0.81,   1.73,   2.08 &  75.76,  94.15,  84.96, 104.01, 116.09  & 6.03e-02,  8.06e-01  &  1.30e+00, 2.55e-01 & NV \\
         &(R-I) &  74 &   1.28,   0.32,   0.80,   1.73,   2.08 & 209.11,  91.13, 150.12, 104.01, 116.09  & 5.36e-02,  8.17e-01  &  1.02e+01, 1.71e-03 & NV \\\hline
\end{tabular}}\\
\label{tab:IDVtab}
(This table is available in its entirety in a machine-readable form in the online journal. A portion is
shown here for guidance regarding its form and content)
\end{table*}

\noindent
Out of $\sim$175 observing nights during the campaign, we have many nights  when multiple image 
frames were observed in any specific optical band. But to study the optical flux and color variability
properties on IDV timescales, we selected only nights with a minimum of ten observations in an optical 
band by a telescope on a particular observing night, and for plotting the IDV LCs, we decided on a minimum 
of twenty observations in an optical band by a telescope on a particular observing night. Using these 
criteria, eleven observing nights qualified for IDV flux and color variability analysis and the results 
are reported in Table 3, while nine multi-band optical IDV flux LCs are plotted in Fig.\ 2. \\ 
\\
To investigate the flux and color variability on IDV timescales on the above nights, we have used 
{\it F}-test, $\chi^{2}$-test, and Levene test analyses which were briefly explained in sections 
3.1.1, 3.1.2, 3.1.3, respectively. Using these tests, the  presence or absence of IDV are reported 
in Table 3 where NV, and Var represent non-variable, and variable natures of LC. It is clearly seen 
from the plots in Fig. 2 as well as the results reported in Table 3 that, perhaps surprisingly, no 
genuine IDV was detected in any of the V, R, and I bands IDV LCs in the nine nights of intensive 
observations that were taken during 20 February 2017 -- 3 April 2017. Given the lack of variability 
in any bands, it is obvious that no color variations were seen on any of these nights on IDV timescales. 
However, it should be noted that nearly all of these nightly observations were relatively short, spanning 
between $\sim$2.5 and $\sim$4 hours, and so the chances of detecting IDV were limited. \\
\\
From the LCs plotted in Figs.\ 1 and 2, it is can be seen that the blazar OJ 287 was in a fairly 
bright state during much of these observations. The brightest state detected in the blazar in the outburst 
in December 2015 was $\sim$13.4 mag in V, 13.0 mag in R, and 12.4 mag in I band \citep{2017MNRAS.465.4423G}. 
Just from these IDV LCs, we detected OJ 287 in the brightest state on 25 February 2017 when the magnitudes 
were $\sim$13.9 mag in V, $\sim$13.6 mag in R, and $\sim$13.0 mag in I.         

\subsection{Short and Long Term Variability}

\subsubsection{Flux Variability}

\noindent
Significant flux variability of OJ 287 on STV and LTV timescales is evident from the four lower panels 
of Fig.\ 1 where the B, V, R, and I band LCs are shown. We have calculated the variability 
amplitude in B, V, R, and I optical bands and these results are reported in Table 4. Observations in the B band were
only carried out using telescopes in Bulgaria and Serbia for a total of sixteen
observing nights between 3 January to 3 April 2017, while we have dense observations
carried out in V, R, and I bands. We noticed that the faintest level of the blazar in B, V, R, I, bands respectively
were 14.921 mag at JD 2457846.39561, 16.164 mag at JD 2458075.58133, 15.670 mag at JD 2458073.96976, and 
14.154 mag at JD 2457866.00547. Similarly, the brightest levels we observed of the blazar in the B, V, R, I, 
bands were 14.126 mag at JD 2457756.53125, 13.582 mag at JD 2457672.36782, 13.179 mag at JD 2457679.620, 
and 12.720 mag at JD 2457691.29261, respectively. The amplitudes of variation in B, V, R, I 
bands are 79.5\%, 258.2\%, 249.1\%, 143.4\%, respectively, but the smaller values for the B and I bands are explained 
by the relative paucity of data for them, as all colors are seen to vary together when data were taken for all of them.
It can be seen from Fig.\ 1 that small flares are superimposed on a long term trend. The implications of these
results are discussed in Section 4.

\begin{table}
\noindent
{\bf Table 4.} Results of STV and LTV flux variations. \\
\\
\centering
\begin{tabular}{cccc} \hline
Band   & Duration                 & Variable  &  A (percent)    \\ 
       & yyyy mm dd -- yyyy mm dd &           &                 \\ \hline
B      & 2017 01 03 -- 2017 04 26 & Var       & ~~79.5  \\
V      & 2016 09 24 -- 2017 12 17 & Var       & 258.2  \\
R      & 2016 09 24 -- 2017 12 16 & Var       & 249.1  \\
I      & 2016 10 09 -- 2017 05 24 & Var       & 143.4  \\\hline
\end{tabular}

Note: Var: Variable; NV: non-variable
\label{tab:flxVarTab}
\end{table} 

\begin{figure}
\centering
\epsfig{figure= 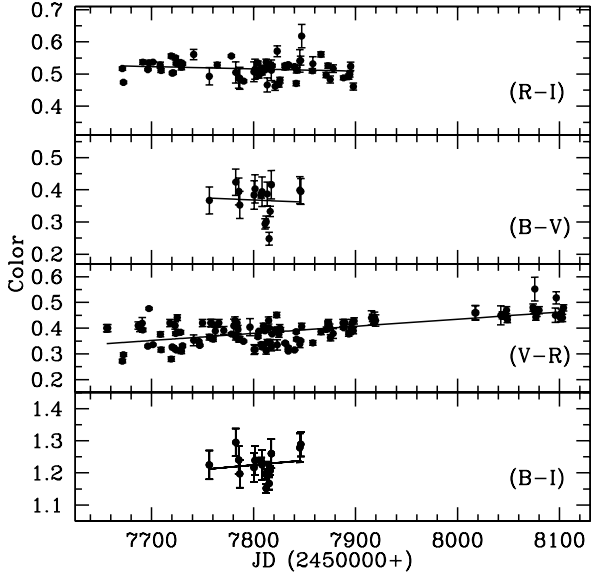,height=3.4in,width=3.4in,angle=0}
\caption{Optical color variability light curves covering the entire monitoring period of OJ 287.}
\label{fig:corVar}
\end{figure}

\subsubsection{Color Variability}

\begin{table}
\noindent
{\bf Table 5.} Color variation with respect to time on short and long timescales. \\
\\
\centering
\begin{tabular}{ccccc} \hline
Color Indices     &  $m_1^a$  &  $c_1^a$  &   $r_1^a$  & $p_1^a$    \\ \hline
R--I & $-$7.011E$-$05 & ~~~1.063 & $-$0.153 & 2.287E$-$01 \\
B--V & $-$1.293E$-$04 & ~~~1.377 & $-$0.059 & 8.296E$-$01 \\
V--R & ~~~2.756E$-$04 & $-$1.770 & ~~~0.634 & 1.029E$-$14 \\
B--I & ~~~2.851E$-$04 & $-$0.999 & ~~~0.156 & 2.888E$-$01 \\\hline
\end{tabular}
\label{tab:corVar}

$^a$ $m_1 =$ slope and $c_1 =$ intercept of CI against JD; \\
$r_1 =$ Pearson coefficient; $p_1 =$ null hypothesis probability \\
\end{table} 

\begin{table}
\noindent
{\bf Table 6.} Color-magnitude dependencies and color-magnitude correlation coefficients on short and long timescales. \\
\\
\centering
\begin{tabular}{ccccc} \hline
Color Indices     &  $m_2^a$  &  $c_2^a$  &   $r_2^a$  & $p_2^a$    \\ \hline
R--I & 7.447E$-$02 & $-$0.513 & 0.487 & 3.319E$-$08 \\
B--V & 2.446E$-$02 & ~~~0.022 & 0.097 & 0.721 \\
V--R & 4.093E$-$02 & $-$0.209 & 0.440 & 1.577E$-$09 \\
B--I & 9.934E$-$02 & $-$0.174 & 0.477 & 0.061 \\\hline
\end{tabular}
\label{tab:CorMag}

$^a$ $m_2 =$ slope and $c_2 =$ intercept of CI against V; \\
$r_2 =$ Pearson coefficient; $p_2 =$ null hypothesis probability \\
\end{table}

\noindent
Optical color variations with respect to time (color vs time) and with respect to V band magnitude
(color vs magnitude) are plotted in Fig.\ 3 and Fig.\ 4, respectively. On visual inspection both the
figures show clear color variations. However, it seems there are no consistent systematic trends in the color variations,
as shown by straight line fits to the color vs time plot in Fig. 3.  However, the color vs magnitude plots in 
Fig.\ 4 show  bluer-when-brighter (BWB) trends, which are significant for the more frequently
measured V--R and R--I colors. Here lines, $Y = mX + c$, are fitted in to 
each panel in Fig.\ 3 and Fig.\ 4. The values of the slopes, $m$, the intercepts, $c$, the linear Pearson 
correlation coefficients, $r$, and the corresponding null hypothesis probability, $p$, results for color 
vs time and color vs magnitude are reported in Table 5 and Table 6, respectively. 

\begin{figure}
\centering
\epsfig{figure= 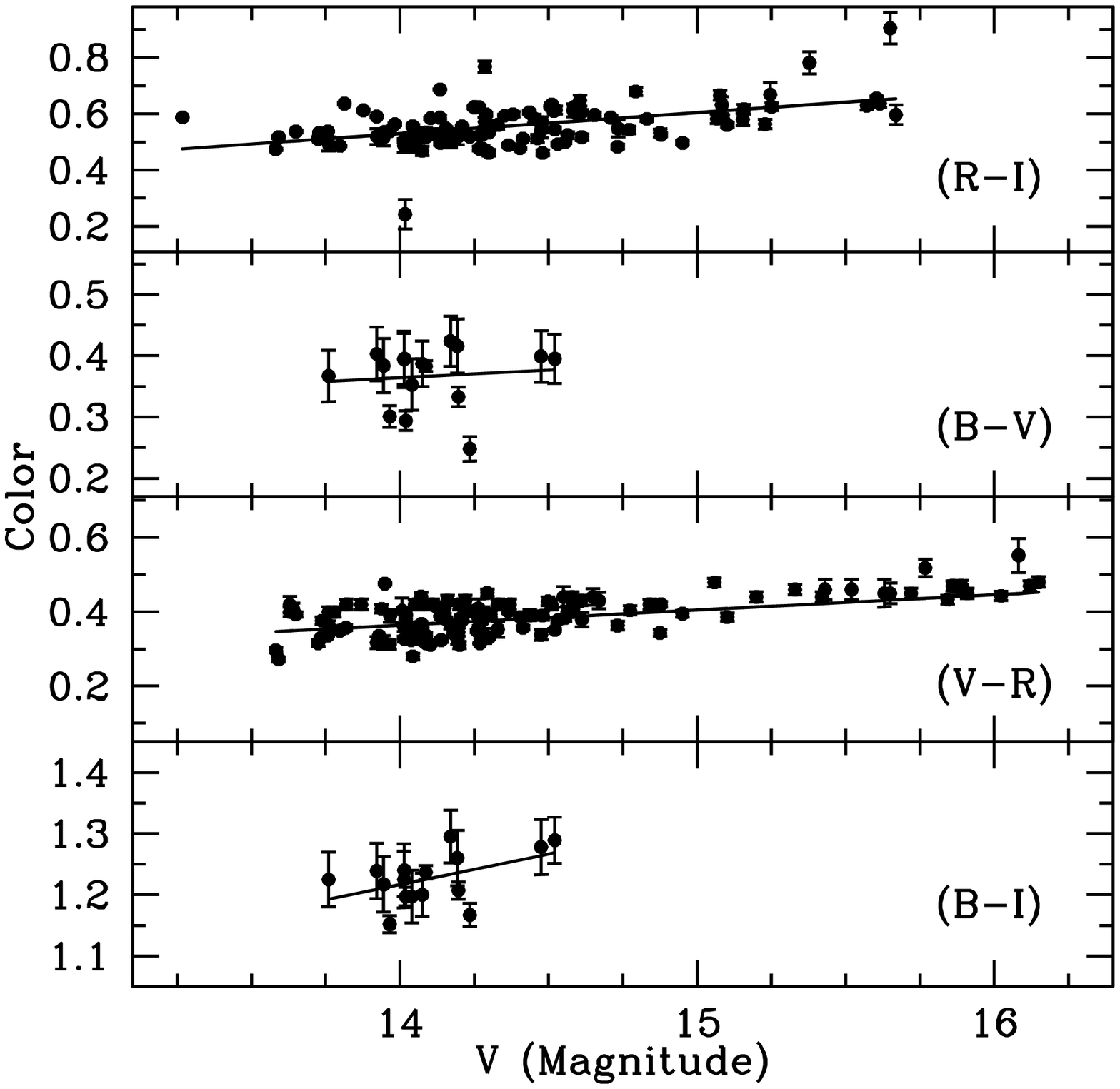,height=3.4in,width=3.4in,angle=0}
\caption{Optical color-magnitude plots of OJ 287 during our entire monitoring period.}
\label{fig:CorMag}
\end{figure}

\subsubsection{Polarization Variability}

\noindent
In Fig.\ 1 we plotted optical magnitudes of OJ 287 as well as degree of polarization and polarization angle.
It is clear from visual inspection of the figure that the source exhibited large variations in 
PD and PA as well as overall flux. We noticed the following combinations of 
variation of flux, degree of polarization, and polarization angle: (i) at
$\sim$ JD 2457682 the flare peak at R = 12.957 mag  corresponds to a PD of 20\%, and 
PA = 15$^\circ$; (ii) at $\sim$ JD 2457755 the flare peak at R = 13.35 is anti-correlated with 
PD at only 9\% and the PA = $-$35$^\circ$; (iii) at $\sim$ JD 2457790 the 
flare peak at R = 13.437 mag  corresponds to PD = 13\% and with PA =
$-$63$^\circ$; (iv)  the lowest flux state at $\sim$ JD 2458074, with R = 15.670 mag, is anti-correlated, 
having a high PD $\sim$ 20\%, and the PA = $-$25$^\circ$. \\

\noindent
Fig.\ 5 presents the polarization in terms of the normalized Stokes parameters, Q/I vs.\ U/I, for these data. 
Here we have used the data taken during JD 2457633 to 2457920, before the $\sim$100 day gap 
when the blazar could not be observed from the ground.  The more limited 
polarimetric observations taken afterwards  have not been used in the analysis. We note there appears 
to be a systematic change in the polarization fluxes with the polarization angle during the 
first $\sim$100 days which evinces a clockwise loop-like structure (shown by black to purple points). During the 
middle part of the observations (red to orange points) there is relatively less change in the Stokes parameters 
with roughly random variations.  However, there is a hint of the beginning of a systematic change over the last 
50 days of the observations (yellow points). Over all, the intensity variations in the Q,U--plane are mostly 
reminiscent of a random-walk, indicating that emission is resulting from different regions with different magnetic 
field orientations \citep{1982ApJ...260..415M}, throughout the course of observations presented in this study.
  
\subsubsection{Spectral Index Variation and Spectral Energy Distribution}

\noindent
We have dense sampling in V and R bands during our whole observing campaign, so we also calculated 
spectral indices for for all the epochs where we have V and R bands data on same JD, following
\citet{2015A&A...573A..69W},

\begin{equation} 
\alpha_{VR} = {0.4\, (V-R) \over \log(\nu_V / \nu_R)} \, 
\end{equation}

\noindent
where $\nu_V$ and $\nu_R$  are effective frequencies of V and R bands, respectively \citep{1998A&A...333..231B}. 
It should be noted that spectral index estimated using this expression differs 
from the usual index (in $\rm F_\nu \propto \nu^{-\alpha}$) by a constant factor
related to the zero-point magnitude of the two bands. The spectral index with
respect to time,  and with respect to V band magnitude are plotted in the bottom 
and top panels of Fig.\ 6, respectively. It is clearly seen that there are large variations in $\alpha_{VR}$ 
between $\sim$1.5 to $\sim$3.2. From Fig.\ 6 (bottom panel) it is clear that the spectral index systematically increased with 
respect to time, and from Fig.\ 6 (top panel), the spectral index increases with respect to increasing V band 
magnitude, confirming the BWB result. The straight line fitting parameters are given in Table 7. \\ 
\\
To further explore the systematic change of the spectral index with time, as well as
magnitude, as seen in Fig.\ \ref{fig:IndxLc} we have generated optical SEDs during different
optical flux states of the source. We have taken quasi-simultaneous B, V, R,
and I band data points at four different states to produce 
optical SEDs. These states are: during outburst 1 from JD 2457756.5 -- 2457757.5; during outburst 2 from
JD 2457811.2 -- 2457812.5; an intermediate state from JD 2457844.5 -- 2457845.5; and a low state from 
JD 2457867.5 -- 2457868.5.  Unfortunately we only have data in two bands (V and R) at the lowest flux 
state of the source on JD 2458074 during this observing campaign and so could not plot an optical SED for it. 
To generate the SEDs, calibrated magnitudes of OJ 287 in B, V, R, I bands are adjusted for  Galactic 
absorption, with A$_{B} =$ 0.102 mag, A$_{V} =$ 0.075 mag, A$_{R} =$ 0.063 mag, and A$_{I} =$ 0.045 mag, 
respectively \citep{1989ApJ...345..245C,1998A&A...333..231B}. 
The SEDs of these four different flux states of the source are plotted in Fig.\ 7. During the observing 
campaign there were other periods of strong flaring as well as intermediate or low flux states of the 
source, but unfortunately we do not have quasi-simultaneous (within a day)  observations in at least three 
optical bands, so those are not considered for making SEDs. \\ 
\\
The most striking observation about the optical SEDs during the current period are their relative 
flatness in the intermediate and low states while there is a slightly rising trend with frequency during 
outbursts 1 and 2. These are very different compared to the previous optical SEDs of the source so far, 
which showed a clear declining trend at higher frequencies
\citep{2018MNRAS.473.1145K,2017MNRAS.465.4423G,2013MNRAS.433.2380K}. The corresponding
spectral indices for the four SEDs with respect to V band magnitude, following Eqn.\ (4), 
are shown in Figure \ref{fig:VRindx}. The indices clearly reflect the variation
between the two bands, with indices being lower for harder spectra, as can be seen
clearly in the SEDs for the respective states.

\section{Discussion}

\noindent
OJ 287 is currently in an enhanced activity phase that started in November
2015 \citep{2018MNRAS.473.1145K,2017MNRAS.465.4423G,2018MNRAS.479.1672K}. The timing
of the beginning of this optical enhancement was 
in accoradance with the inspiraling binary SMBH model
\citep[and references therein]{2016ApJ...819L..37V} which attributes the $\sim12$
years quasi-periodicity seen in the optical data to the impact of the secondary
SMBH on the accretion disk of the primary. Apart from confirming the model predictions,
studies and observations of OJ 287 since then have resulted in reporting of many
new features in the spectral, temporal, and polarization domains. These include
a possible thermal bump in NIR-optical region consistent with primary SMBH accretion
disk emission \citep{2018MNRAS.473.1145K}, a change in shape of $\gamma$-ray SEDs
and shifts in its peak position \citep{2018MNRAS.473.1145K,2018MNRAS.479.1672K},
and the first ever detection at very high energies \citep[VHEs, E $>$ 100 GeV;]
[]{2017arXiv170802160O}. The work presented here is part of ongoing effort to explore
the source activity and features between the two claimed disk impacts via dense
follow-up at optical energies. A significant part of the present data has already
been presented in another work focusing on MW aspects of the source \citep{2018MNRAS.479.1672K},
but that paper was concerned with timescales of interest associated with the MW data
cadence. Here, we are focused on investigation on diverse timescales from IDV to LTV
which are allowed by the current data and which was not possible with the MW data either
due to the low cadence of X-ray and $\gamma$-ray data and/or the lack of good
photon statistics for the latter on all the timescales of the optical data considered here. \\

\begin{figure}
\centering
\epsfig{figure= 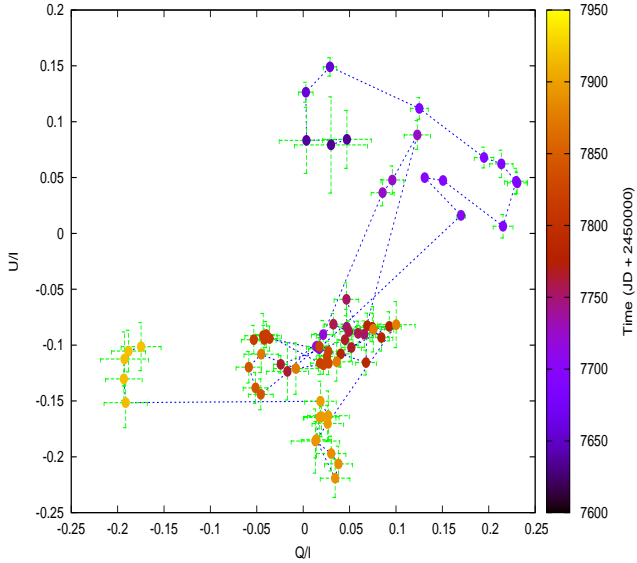,height=3.3in,width=3.0in,angle=-90}
\caption{Fractional polarization variations over the course of data presented in this study.}
\label{fig:Stoke}
\end{figure}

\begin{figure}
\centering
\epsfig{figure= 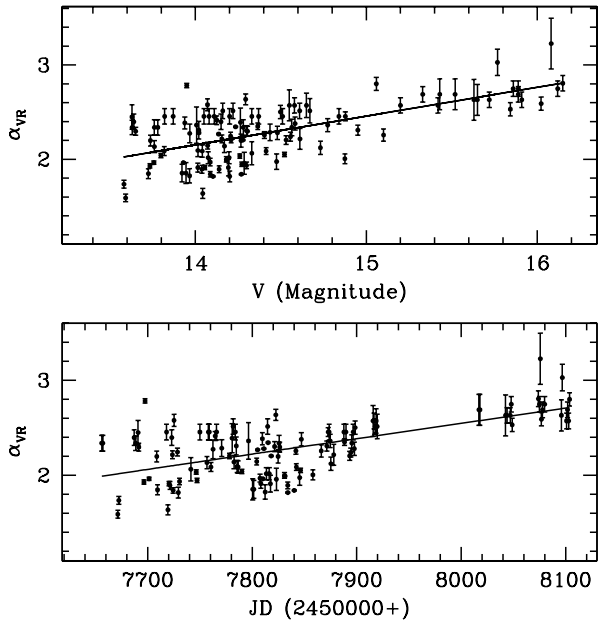,height=3.4in,width=3.4in,angle=0}
\caption{Optical spectral index variation with respect to time and V band magnitude covering the entire 
monitoring period of OJ 287.}
\label{fig:IndxLc}
\end{figure}

\begin{figure}
\centering
\epsfig{figure= 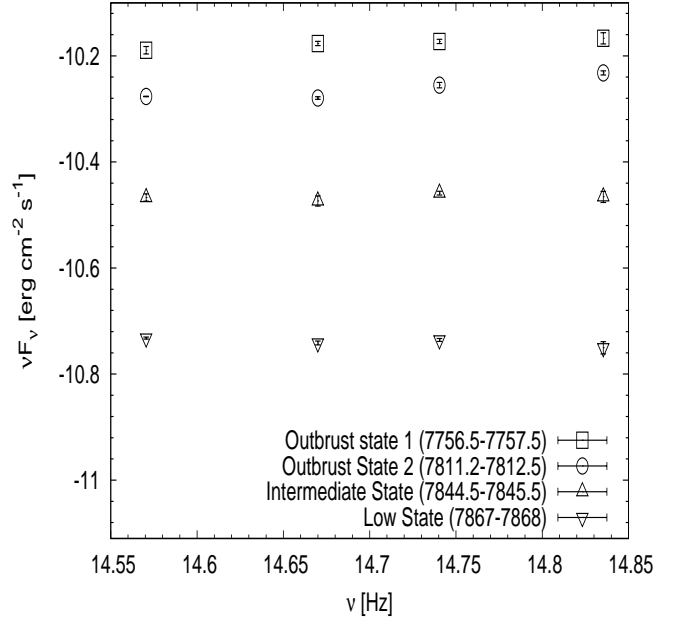,height=3.4in,width=3.4in,angle=0}
\caption{Optical spectral energy distribution (SED) of OJ 287 at different flux state.}
\label{fig:SEDs}
\end{figure}

\noindent
The data presented here correspond to the September 2016 -- December 2017 period when
we have obtained extensive optical photometric and polarimetric monitoring of OJ 287 on a
total of $\sim$175 nights from seven different optical telescopes around the globe.
 Archival data from Steward observatory are also included in our studies. 
We have searched for flux and color variations on IDV, STV and LTV timescales.
We have also searched for color variations with respect to time, color
dependence on magnitude, spectral index variation, SED changes, and polarization
variations on LTV timescales. Investigation of IDV on 11 nights when we have
quasi-simultaneous multi-band (V, R, I) observations showed no significant IDV in fluxes
or colors (Table \ref{tab:IDVtab}). On the other hand, on STV and
LTV timescales the light curves
show strong evidence of major flux changes, with amplitude variations reaching $\sim$250\%
(Table \ref{tab:flxVarTab}), including multiple instances of flaring. The lower detected variabilities for 
the B and I bands are almost certainly due to the lesser amounts of 
data for them. There is some evolution of the best measured (V$-$R) color  with time (Table \ref{tab:corVar}, Fig.\
\ref{fig:corVar}) but there are clear systematic variations of color with respect to the V-band
measurements (Table \ref{tab:CorMag}, Fig. \ref{fig:CorMag}), referred to generally
as a BWB trend which is a result of a new non-thermal high-frequency peaked blazar (HBL) 
component \citep{2018MNRAS.479.1672K}. \\
\\
In the polarization domain, the source reflects the activity seen in the flux variations on similar timescales. 
Broadly, most of the flux increments are accompanied by an increase in
PD which are also associated with frequent changes in the PA by amounts of $\lesssim 50^\circ$.  The Stokes 
parameters show a systematic clockwise trend during the first hundred days, followed by an erratic
variation and finally a return to a systematic trend towards the end (Fig. \ref{fig:Stoke}), indicating the
importance of magnetic field changes. We note that the
transition from the clockwise trend to the erratic one coincided with the VHE detection and the return
to a trend occurred when the source was no longer detectable at those energies \citep{2017arXiv170802160O}.  It should
be further noted that PA (after correcting for the $\pm 180^\circ$ ambiguity) shows
a smooth systematic change of $\sim 125^\circ - 150^\circ$ with small
amplitude variations superimposed on it. \\
\begin{figure}
\centering
\epsfig{figure= 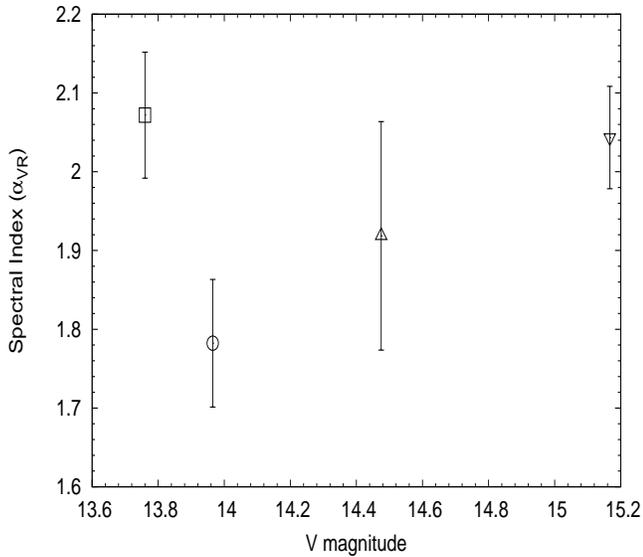,height=3.0in,width=3.4in,angle=0}
\caption{Variation of optical spectral index of OJ 287 with V magnitude  during two outburst states, 
an intermediate state, 
and a low state, where the symbols have the same meaning as in Fig.\ 7.}
\label{fig:VRindx}
\end{figure}

\noindent
The trends and variations seen in the temporal and polarization domains are also
reflected in the spectral domain where the VR spectral index shows a systematic
change with both V-band magnitude as well as time (Fig. \ref{fig:IndxLc}). This
similar trend with both V-band measurement and flux is reflective of a systematic
decline of emission level with time, as can be seen in the light curves. Interestingly,
the declining trend is similar to the systematic trend of change in the PA. Further,
the flat or uprising optical SEDs suggest that this trend is due to broadband emission.
Though there is strong variability with some systematic trends, the variations during
the current phase are very different from those seen during our
previous observation campaign \citep{2017MNRAS.465.4423G}. Most interestingly, the current data  
do not display IDV variability
despite similar large amplitude variations on LTV and STV timescales, thereby suggesting that the observed
variability is governed by regions of larger sizes corresponding to the LTV/STV timescales. \\
\\
Blazar flux variability on IDV timescales is the most puzzling, and during low states may allow us to probe very 
close to the central  SMBH. IDV in high flux states can be due to evolution of the electron energy density 
distribution of the relativistic charged particles in which shocks will accelerate  relativistic particles in  
turbulent regions of plasma jets which then cool and lead to a variable synchrotron emission 
\citep{1992vob..conf...85M,2014ApJ...780...87M,2015JApA...36..255C,2017ApJ...843...81O}.
The most extreme IDV might require acceleration of small regions within the jets to extremely high Lorentz 
factors \citep[e.g.][]{2009MNRAS.395L..29G}. Optical flux variability detected on IDV timescales in low-states 
can be explained by models based on the accretion disk \citep[e.g.,][]{1993ApJ...406..420M,1993ApJ...411..602C}.
Our lack of detection of genuine IDV in any of the 11 densely sampled nights for OJ 287 indicates that during 
this period the jet emission was quite uniform and that relativistic shock directions did not quickly 
change with respect to our line of sight. Here we can safely rule out accretion disc based models because 
source was observed in an overall high flux state, when jet emission must dominate and no flux or color 
variations were noticed on IDV timescales. \\
\\
Blazar emission on STV and LTV timescales are dominated by non-thermal jet emission throughout the EM spectrum 
and can also explain the optical flux and polarization variability on diverse timescales. Shock-in-jet models 
\citep[e.g.][and  references therein]{1985ApJ...298..301H,1985ApJ...298..114M,2001MNRAS.325.1559S,2008ApJ...689...68G,
2011ApJ...727...21J} can explain the general behaviour of flux variability on diverse timescales, while the 
polarization variability also can be explained by these models 
\citep[e.g.][and the references therein]{2008Natur.452..966M,2013ApJ...768...40L} 
particularly when supplemented with turbulence \citep{2014ApJ...780...87M}. 
Changes in the physical parameters set up close to  the base of the jet including velocity, electron density, magnetic
field, etc., can  produce a new shock which can lead to a flaring event when moving along the inhomogeneous
relativistic jet. Geometrical effects from jet bending,  precession or internal helical structures  can lead to changes in the
Doppler boosting of the jet emission which can produce a wide variety of flux variations on STV and LTV timescales in blazars
\citep[e.g.][]{1992A&A...255...59C,1992A&A...259..109G,2016ApJ...820...12P}. \\
\\
The four optical SEDs (Fig.\ \ref{fig:SEDs}) during different flux
states are almost flat, except for changes in the overall level of emission. In fact, they suggest
more emission at the high frequency end, which is very different compared to previous SEDs
\citep[and references therein]{2018MNRAS.473.1145K,2017MNRAS.465.4423G,2013MNRAS.433.2380K} where normally
declining emission is seen. Since any thermal features are not expected to  vary appreciably on timescales
of days or months, the flatness of the SED and the blue bump emission possibly
seen in our previous work \citet[see also \citet{2018MNRAS.473.1145K}]{2017MNRAS.465.4423G}
suggest a contribution from another component in the optical that makes the emission increase at high 
frequencies \citep[e.g.][]{2018MNRAS.479.1672K}. This is clear from the investigation of the broadband 
SED by \citet{2018MNRAS.479.1672K} with OJ 287 SEDs being a sum of LBL, which is the typical SED of OJ 287, 
and a new HBL spectral component during the high activity states. This new component, peaking in the UV-X-ray 
region, is responsible for the relative flatness of the SEDs. This is also consistent with the high PD and 
strong changes of it during these observations, often associated with PA swings and the systematic variation 
in the fractional polarization suggest the second component to be non-thermal in nature, as also reflected 
in the strong variability of the optical V-R spectral index on daily timescales (Fig.\ \ref{fig:IndxLc}). 
Although OJ 287 is fundamentally a BL Lac object and has at most very weak emission lines,
they may be present at the level of $\sim 10^{42-43} {\rm erg s}^{-1}$, and were apparently seen
during the interaction time suggested by the binary SMBH model \citep{2010A&A...516A..60N} and also
might be relevant for explaining the overall MW SED shift in 
the $\gamma$-ray peak \citep{2017MNRAS.465.4423G, 2018MNRAS.473.1145K}. \\
\\
In our flux and polarization monitoring campaign of OJ 287 during 2016 -- 2017, we
noticed interesting relations between the fluxes, degree of polarization, and
polarization angle. There is a systematic swing of $\sim$150$^{\circ}$ in PA
 from $\sim$ JD 2457630 -- JD 2457850, with a few superimposed short term
fluctuations of up to $\sim$50$^{\circ}$. But during this period the degree of
polarization has large variations, from a few percent to over 20 percent. These
changes in flux, PD, and PA are quite complex. 
Interestingly, the fractional polarization shows a systematic clockwise variation
during the first 100 days, followed by an essentially random trend and again returning
to systematic variation towards the end. It should be noted that the lack of observations
before MJD $\sim$ 57650 allow an ambiguity of $\pm 180^{\circ}$ in representation
of PA. In our presentation of the PA changes here we chose a smooth variation
over a big jump, and so it differs by $180^{\circ}$ to the presentation of some similar 
data \citet{2017Galax...5...83V}.
The choice is based on the fact the PD variation seen here is almost always associated
with PA change and the fact that the fractional polarization shows both systematic and
chaotic trends. Further, in random variation models, a sudden jump of no more than $\pm 90^{\circ}$ is 
expected \citep[e.g.][]{2014ApJ...780...87M}. \\
\\
In the basic shock-in-jet 
model, where the shocked region strengthens the ordering of the magnetic field, one 
can expect a positive correlation 
between flux and polarization, i.e., an increase in polarization with an increase in flux
\citep{1985ApJ...298..114M,1996ASPC..110..248M,2008ApJ...672...40H}. 
There are several cases in which blazar flux and degree of polarization show such positive correlations 
\citep[e.g.][and references therein]{2008A&A...492..389L,2013ApJ...768...40L,2016MNRAS.461.3047L}.
Detection of anti-correlated flux and degree of polarization is rare but has been occasionally noticed before 
\citep[e.g.][]{2014ApJ...781L...4G}. During this observing campaign, we noticed that when 
the source goes into the lowest flux state, at $\sim$ JD 2457865, the PD
is rather high and there is evidence of large swings $\sim$ 70$^{\circ}$ of the PA. \citet{2008Natur.452..966M} gave a generalized model for variation in optical flux, degree 
of polarization, and polarization angle. The model involves a shock wave leaving the close vicinity 
of the central SMBH and propagating down only a portion of the jet's cross section which 
leads to the disturbance following a spiral path in a jet that is accelerating and becoming 
more collimated. \citet{2013ApJ...768...40L} extended the work of \citep{2008Natur.452..966M}  and applied 
this generalized model to  multiwavelength variations of an outburst detected in the 
blazar S5 0716+714. In the model of \citet{2013ApJ...768...40L},
if one changes the bulk Lorentz factor $\Gamma$, even if the remaining parameters 
(e.g. jet viewing angle, temporal evolution of the outburst, shocked plasma compression ratio $k$, spectral
index $\alpha$, and pitch angle of the spiral motion) are kept constant, different combinations in the variations in flux,
polarization, and polarization angle can be observed. 

\section{Summary}

\noindent
We summarize below our results: \\
\\
$\bullet$ The blazar OJ 287 was in a fairly bright state between September 2016 and December 2017
and several large and small flares were observed in optical bands. \\
$\bullet$ Using our selection criteria, we had eleven nights during which multi-band intra-day LCs could be
extracted but we never saw fast (IDV) variations in flux or color. \\ 
$\bullet$ On longer STV and LTV timescales OJ 287 showed large amplitude flux variation in all B, V, R,
and I bands with variability in respective band similar to what was found
in the previous study \citep{2017MNRAS.465.4423G}. \\ 
$\bullet$ Color variations are noticed on STV and LTV timescales in both color vs time and color
vs magnitude plots. A bluer-when-brighter trend is noticed
between the best sampled V and R bands. \\ 
$\bullet$ There are strong variations in degree of polarization and large swings in polarization angle. 
For most of the time, both flux and polarization show complex variations. \\
$\bullet$ On two occasions around  JD 2457755 and JD 2458074, we noticed that there are strong 
evidences of anti-correlation in flux with degree of polarization and polarization angle. \\
$\bullet$ Through plotting  the Stokes parameters, we observed that the fractional polarization exhibited 
a systematic clockwise trend with time during the first hundred days, followed by a more restricted and 
essentially random variation, and then it appears to revert to a systematic variation.
This duration and trend are coincident with the source's VHE activity
\citep{2017arXiv170802160O}, suggesting a role magnetic field for that activity.

\section*{ACKNOWLEDGMENTS}

\noindent
Data from the Steward Observatory
spectropolarimetric monitoring project were used. This program is supported by Fermi Guest
Investigator grants NNX08AW56G, NNX09AU10G, NNX12AO93G, and NNX15AU81G. \\
\\
We thank the anonymous referees for useful comments.
The work of ACG and AP are partially supported by Indo-Poland project No. DST/INT/POL/P–19/2016 
funded by Department of Science and Technology (DST), Government of India. ACG's work is also 
partially supported by Chinese Academy of Sciences (CAS) President's International Fellowship Initiative 
(PIFI) grant no.\ 2016VMB073. HG acknowledges financial support from the Department of Science 
\& Technology, India through INSPIRE faculty award IFA17-PH197 at ARIES, Nainital. PJW is 
grateful for hospitality at KIPAC, Stanford University, and SHAO during a sabbatical. PK 
acknowledges support from FAPESP grant no.\ 2015/13933-0. The Abastumani team acknowledges 
financial support by the Shota Rustaveli National Science Foundation under contract FR/217554/16. 
OMK acknowledges China NSF grants NSFC11733001 and NSFCU1531245. SMH’s work is supported by the 
National Natural Science Foundation of China under grant No. 11203016, Natural Science Foundation 
of Shandong province (No. JQ201702), and also partly supported by Young Scholars Program of 
Shandong University, Weihai. The work of ES, AS, RB was partially supported by the Bulgarian 
National Science Fund of the Ministry of Education
and Science under the grants DN 08-1/2016 and DN 18-13/2017. GD and OV gratefully acknowledge the
observing grant support from the Institute of Astronomy and NAO Rozhen, BAS, via bilateral joint
research project ``Study of ICRF radio-sources and fast variable astronomical objects" (the head
is GD). This work is a part of the Projects no.\ 176011 ``Dynamics and kinematics of celestial
bodies and systems", no.\ 176004 ``Stellar physics" and no.\ 176021 ``Visible and invisible matter
in nearby galaxies: theory and observations" supported by the Ministry of Education, Science and
Technological Development of the Republic of Serbia. AG acknowledges full support from the Polish 
National Science Centre (NCN) through the grant 2012/04/A/ST9/00083 and partial support from the 
UMO-2016/22/E/ST9/00061 . {\L}S is supported by Polish NSC grant UMO-2016/22/E/ST9/00061. 
MFG is supported by the National Science Foundation of China (grants 11473054 and U1531245). 
ZZ is thankful for support from the CAS Hundred-Talented program (Y787081009).

\begin{table}
\noindent
{\bf Table 7.} Spectral index variation with respect to JD and V band magnitude for entire period of 
observation campaign of OJ 287. \\
\\
\centering
\begin{tabular}{lcccc} \hline

Parameter  &  $m_3^a$  &  $c_3^a$  &   $r_3^a$  & $p_3^a$    \\ \hline
$\alpha_{VR}$ vs JD      & 1.612E$-$03 & $-$10.350 & 0.634 & 1.013E$-$14 \\
$\alpha_{VR}$ vs V (mag) & 3.054E$-$01 & $-$~2.123 & 0.650 & 1.235E$-$15  \\\hline
\end{tabular}

$^a$ $m_3 =$ slope and $c_3 =$ intercept of $\alpha$ against JD or V; \\
$r_3 =$ Pearson coefficient; $p_3 =$ null hypothesis probability \\
\end{table}

\clearpage

\end{document}